\documentclass[aps,prd,twocolumn,a4paper,superscriptaddress,10pt,floatfix]{revtex4-2}
\usepackage{graphicx,multirow,latexsym}
\usepackage{hyperref}
\usepackage[british]{babel}
\usepackage{amsmath,amsmath}
\usepackage{array}
\usepackage{tabularray}
\usepackage{soul}

\begin{document}
\newcommand{\be}{\begin{equation}}
\newcommand{\ee}{\end{equation}}
\newcommand{\bq}{\begin{eqnarray}}
\newcommand{\eq}{\end{eqnarray}}
\newcommand{\bsq}{\begin{subequations}}
\newcommand{\esq}{\end{subequations}}
\newcommand{\bc}{\begin{center}}
\newcommand{\ec}{\end{center}}
\newcommand{\vos}[1]{{\left<#1\right>}}
\newcommand{\Hub}{\mathcal{H}}
\newcommand{\dd}[1]{\text{d}{#1}}
\renewcommand{\vec}[1]{\bf{#1}}
\newcommand{\matr}[1]{\boldsymbol{#1}}
\newcommand{\dpartial}[2]{\frac{\partial {#1}}{\partial {#2}}}
\newcommand{\ddfrac}[2]{\frac{\dd{#1}}{\dd{#2}}}
\newcommand{\ddpartial}[2]{\frac{\partial^2 {#1}}{\partial {#2}^2}}
\newcommand{\dddfrac}[2]{\frac{\dd{}^2 {#1}}{\dd {#2}^2}}
\newcommand{\dprime}[1]{{#1^\prime}}
\newcommand{\ddprime}[1]{{#1^{\prime\prime}}}

\title{Scaling solutions for current-carrying cosmic string networks. II. Biased solutions}
\author{F. C. N. Q. Pimenta}
\email{up200908672@edu.fc.up.pt}
\affiliation{Centro de Astrof\'{\i}sica da Universidade do Porto, Rua das Estrelas, 4150-762 Porto, Portugal}
\author{C. J. A. P. Martins}
\email{Carlos.Martins@astro.up.pt}
\affiliation{Centro de Astrof\'{\i}sica da Universidade do Porto, Rua das Estrelas, 4150-762 Porto, Portugal}
\affiliation{Instituto de Astrof\'{\i}sica e Ci\^encias do Espa\c co, Universidade do Porto, Rua das Estrelas, 4150-762 Porto, Portugal}
\date{\today}

\begin{abstract}
The charge-velocity-dependent one-scale model is an extension of the canonical velocity-dependent one-scale model which explicitly incorporates additional degrees of freedom on the string worldsheet, such as arbitrary currents and charges, expected in physically realistic models. A previous paper [Pimenta and Martins, Phys. Rev. D 110, 023540 (2024)] started an in-depth classification of its possible asymptotic scaling solutions, with the goal of identifying distinguishing features which may be tested against numerical simulations or future observations. This earlier analysis restricted itself to unbiased solutions; in the present one we relax this assumption, allowing for the possibility of energy loss biases between the bare string and the charge/current, or between the charge and the current themselves. We find additional scaling solutions, some of which are parametric extensions of the unbiased ones while others are new scaling branches which do not exist in the unbiased case. Overall it remains the case that there are three broad classes of solutions, mainly depending on the balance between the expansion rate and the available energy loss mechanisms. Our results enable a quantitative calibration of the model, using forthcoming high-resolution numerical simulations.
\end{abstract}
\maketitle
%%%%%%%%%%%%%%%%%%%%%%%%%%%%%%%%%%%%%%%%%%%%%%%%%%%%%%%%%%%%%%%%%%%%%%%%%
\section{\label{01intro}Introduction}

Cosmic strings have been extensively studied, as the most interesting fossil relic of cosmological phase transitions \cite{Kibble76,VSbook}. Most existing studies, including
those providing constraints from and forecasts of their observational signals, rely on assumptions of featureless networks, neglecting the additional degrees of freedom on the string worldsheet, e.g., charges and currents. This assumption is not at all satisfactory, since such additional degrees of freedom are expected to occur in physically realistic models \cite{Witten_1984,Witten85,Copeland_2010}, and they will impact the evolution of the networks and therefore also their cosmological consequences. For example, the main energy loss mechanism for superconducting strings is expected to be the electromagnetic one, with gravitational radiation being heavily subdominant \cite{EMG1,EMg2}.

The added degrees of freedom make a fully quantitative study of cosmic superstrings a very challenging task, but general current-carrying string networks provide useful proxies for them. Recent years have seen a significant progress in numerically simulating these networks at high resolution \cite{Saffin05,Urrestilla_2008,Lizarraga16,Correia_2022,Battye23,Simulations}. Meanwhile, the canonical velocity-dependent one-scale (VOS) model \cite{VOS1,Martins_2002,Martins_2016} has been extended to allow for generic charges and currents on the string worldsheet, leading to the charge-velocity-dependent one scale (CVOS) model \cite{Martins_2021,Rybak_2023}. Both the VOS and CVOS models include phenomenological free parameters, to be calibrated from numerical simulations; this calibration has been successfully achieved for the plain VOS \cite{Correia21}, but for the CVOS it is work in progress, with preliminary results reported in \cite{Simulations}.

In \cite{PaperI}, hereafter referred to as Paper I, we started to study and classify the possible asymptotic scaling solutions of the CVOS model, exploring the impact of the expansion rate of the universe and other physical mechanisms impacting the network dynamics. This classification is an important preamble to the numerical calibration; it is analogous to the one previously done for chiral strings \cite{Oliveira_2012} and wiggly strings \cite{Almeida_2021,Almeida_2022}. Our focus is on the physical scaling solutions; other mathematical solutions (which are asymptotic solutions of the CVOS dynamical equations but are not physically relevant) have been listed and classified elsewhere \cite{Thesis}. The overarching result of Paper I is that scaling solutions can be divided into three classes, primarily determined by the cosmological expansion rate. For a fast enough expansion rate the charges and currents decay and the network evolves towards the plain Nambu-Goto case. Conversely, for slow expansion rates the charge and current can dominate the network dynamics and prevent the linear scaling behaviour typical of plain networks. In between these two regimes, and for one specific expansion rate, there is a third regime in which the network is in  \emph{full scaling}: an extension of the canonical linear scaling regime which also includes constant values of the charge and current. The specific value of this threshold expansion rate depends on the available energy loss mechanisms and their strength, which in the model are described by suitable model parameters, but under some assumptions it can occur at the matter-dominated era.

Another relevant property of the various scaling solution branches is whether or not they are \emph{chiral}, i.e. whether or not the network's average charge and current are equal. Paper I has presented examples of chiral scaling solutions, but also of non-chiral ones---either because the charge and current have different time dependencies or because, when they have the same time dependence, their ratio is not unity. It has further shown that whether or not each of these solutions can exist is determined, among other factors, by the microphysics of the underlying model. This is a significant result, which we further discuss in what follows.

One limitation of Paper I is that it only addressed \emph{unbiased} scaling solutions, in a sense to be rigorously defined in the next section. Presently we relax this assumption, and explore further scaling solutions which allow for the possibility of energy loss biases between the bare string and the additional degrees of freedom (i.e., the charge and current), and also between the charge and the current themselves. As we will see the overall picture described in the two previous paragraphs still holds. Nevertheless, allowing for biases does impact the energy loss mechanisms, and this leads to generalized scaling solutions, which are parametric extensions of the ones reported in Paper I, but in some cases also to new solutions branches that do not exist in the unbiased case.

The structure of the rest of the paper is analogous to that of Paper I, and is as follows. In Sect. \ref{02VOS} we provide a brief introduction to the CVOS model; unavoidably this overlaps with Sect. II of Paper I (being a shorter version of it), but it is included here for the purposes of making the present work self-contained and of defining the relevant quantities. Solutions in Minkowski spacetime are described in Sect. \ref{03nomechanisms}.  Our main results are in Sects. \ref{03nolosses} and \ref{04losses}, which discuss the possible asymptotic scaling solutions in expanding universes, without and with losses respectively, and the conditions under which each of them may occur. In fact, Sect. \ref{03nolosses} turns out to be a trivial one, for reasons which will become clear therein. Finally, an overall discussion of these solutions and our conclusions, drawing on the results of Paper I as well as the present ones, are presented in Sect. \ref{05conclusions}.

%%%%%%%%%%%%%%%%%%%%%%%%%%%%%%%%%%%%%%%%%%%%%%%%%%%%%%%%%%%%%%%%%%%%%%%%%
\section{\label{02VOS}CVOS Model overview}

One of the physical assumptions of the canonical VOS model is that a single characteristic length scale describes both the network correlation length and the network energy, but this will not apply in the presence of charges or currents on the string worldsheet. The natural extension is to keep the definition of the correlation length scale, $\xi$, as the typical string separation (related to the bare string energy) and defining an additional length scale, $L$, related to the total energy of the network.

The details of this approach for generic charges and currents can be found in \cite{Martins_2021}, and start with the action
\begin{equation}
    S=-\mu_0\int f(\kappa)\sqrt{-\gamma}\dd\sigma^2
\end{equation}
where $\mu_0$ has units of mass squared, and the generic function $f(\kappa)$ depends on the state parameter $\kappa$ defined from a scalar field, $\varphi$, as
\begin{equation}
    \kappa= q^2-j^2
    = \frac{1}{a^2\dprime{\vec{x}}^2}\left(\varepsilon^2\dot{\varphi}^2-\dprime{\varphi}^2\right)
\end{equation}
where the microscopic charge $q^2$ and current $j^2$ have been implicitly defined, $a$ and $\varepsilon$ are the scale factor and coordinate energy per unit length respectively, and the last equality highlights the fact that the chiral limit studied in \cite{Oliveira_2012} corresponds to the limit $\kappa\to 0$. Various specific examples of equations of state $f(\kappa)$ are briefly discussed in \cite{Martins_2021}.

The total energy of the network is
\begin{equation}
     E= a\mu_0\int f \varepsilon \dd\sigma -a\mu_0\int 2q^2\frac{\dd f}{\dd \kappa}\varepsilon\dd\sigma\,,
\end{equation}
where the first term with $f=1$ is the bare string energy and the second includes the energy in the additional degrees of freedom. The macroscopic equation for the total energy is therefore:
\begin{equation}
    \frac{E}{E_0} = \vos{f} - 2\vos{q^2\frac{\dd f}{\dd \kappa}} = F - 2Q^2\dprime{F}\,,
\end{equation}
where $E_0$ is the bare string energy as defined in the VOS model, the variables have been assumed to be uncorrelated to obtain the last equality and capital letters should be understood as the expected average value of their microscopic counterparts, given by
\begin{equation}
    \left<\mathcal{O}\right> \equiv \frac{\int \mathcal{O}\varepsilon {\dd \sigma}}{\int \varepsilon \dd\sigma}
\end{equation}
In particular, the macroscopic charge and current are defined as $Q^2\equiv\left<q^2\right>$ and equivalently for $J^2\equiv\left<j^2\right>$. From these definitions, $F$ and its derivatives encode information on the microscopic equation of motion of the model under consideration. We note that the uncorrelated variables assumption, based on which the CVOS equations are derived, is supported by numerical simulations of Nambu-Goto strings, as discussed in \cite{Martins_2021}. Ideally one would like to further support this assumption, for the specific case of charges and currents, with field theory simulations of current carrying strings. While such simulations are currently being developed, at the time of writing they are not sufficiently developed to carry out such a test, so the uncorrelated variables remains a strong assumption requiring further testing. Finally, one can rewrite this in comoving units and also make explicit the relation between the comoving length scale, $L_c$, and its correlation length counterpart by defining
\begin{equation}
     \xi_c= L_c \sqrt{F-2Q^2\dprime{F}} = W L_c\,,
    \label{eq:lc_xic}
\end{equation}
where the last relation defines the convenient function $W$. This function, being the square root of the ratio of the total and bare energies, is always larger than (or at most equal to) unity.

Under these assumptions, the microscopic equations of motion can be averaged by the usual VOS procedure \cite{Martins_2016} to obtain the following macroscopic equations \cite{Martins_2021}
\begin{subequations}
\label{eq:gvos}
\bq
    \dot{L}_c &=& \Hub L_c\left[v^2-\left(1-v^2\right)\frac{Q^2+J^2}{W^2}\dprime{F}\right] \\
    \dot{v} &=& \left(1-v^2\right)\left[\frac{k_v}{WL_c}\left(1+2\frac{Q^2+J^2}{W^2}\dprime{F}\right)\right]\nonumber\\
    & &-\left(1-v^2\right)\left[2v\Hub\left(1+\frac{Q^2+J^2}{W^2}\dprime{F}\right)\right]\\
    \left(J^2\right){\dot{}} &=& 2J^2\left(\frac{vk_v}{L_cW}-\Hub\right)\\
    \left(Q^2\right){\dot{}}  &=& 2Q^2\frac{\dprime{F}+2J^2\ddprime{F}}{\dprime{F}+2Q^2\ddprime{F}}\left(\frac{vk_v}{L_cW}-\Hub\right) \\
    \dot{\xi}_c &=& \Hub\xi_c v^2+\frac{Q^2+J^2}{W^2}\left(\Hub\xi_c v^2-v k_v\right)\dprime{F}\,;
\eq
\end{subequations}
here dots denote derivatives with respect to conformal time and $k_v$ is the usual VOS momentum parameter \cite{Martins_2002}. Note that the last equation is not independent from the others.

Since they stem from microphysics, these equations do not include energy loss mechanisms beyond the cosmological expansion---the latter being accounted for by the comoving Hubble parameter $\Hub$. Relevant mechanisms include losses due to loop production, possible charge and/or current losses (the two are different nonlinear processes and can occur at different rates), and also bias parameters, describing e.g. whether a region with a high charge or current is more or less likely to be part of loop production events, which would remove it from the long-string network. The phenomenological modelling of these effects is described in detail in \cite{Martins_2021,Linear,Rybak_2023}, and its outcome is the addition of the following terms to the previous equations
 \begin{subequations}
\bq
    \dot{L}_c &=& \dots + \frac{g}{W}\frac{\tilde{c}}{2}v\\
    \left(J^2\right){\dot{}} &=& \dots + \rho\tilde{c}\frac{v}{L_c}\frac{(g-1)W}{\dprime{F}-2Q^2\ddprime{F}}\\
    \left(Q^2\right){\dot{}} &=&\dots + (1-\rho)\tilde{c}\frac{v}{L_c}\frac{(g-1)W}{\dprime{F}+2Q^2\ddprime{F}}\\
    \dot{\xi}_c &=& \dots + \frac{\tilde{c}}{2}v\,,
\eq
\end{subequations}
while the velocity equation is unchanged. Here $\tilde{c}$ is the canonical loop chopping efficiency \cite{VOS1} and is expected to be a constant, while $g$ and $\rho$ are bias parameters and are not necessarily constants. The first of these encodes possible overall biases of the additional degrees of freedom with respect to the bare string, and a detailed analysis in \cite{Rybak_2023} suggests that it can be written in the following generic form
\begin{equation}\label{defg}
    g = 1- g_Q\frac{\dprime{F}+2Q^2\ddprime{F}}{F-2Q^2\dprime{F}}Q^2-g_J\frac{\dprime{F}-2Q^2\ddprime{F}}{F-2Q^2\dprime{F}}J^2\,,
\end{equation}
where $g_Q$ and $g_J$ are dimensionless constants describing how much time-like and space-like components of the network's current are lost to loop production. Here we explicitly see the impact of the model's equation of motion, encoded in $F$ and its first and second derivatives, in determining the amount of bias, if there is any. The second bias parameter, which is only relevant if $g\neq1$, describes biases between charge and current, and should be $\rho=1/2$ in the unbiased case.

The analysis of Paper I only considered the unbiased $g=1$ case; the present work relaxes this assumption. Note that the form of Eq. (\ref{defg}) implies that scaling solutions where charge and current decay will be asymptotically unbiased. On the other hand, for constant charge and/or current solutions one may still expect a constant $g$, but not necessarily an unbiased one. Finally, for growing charges and/or currents the asymptotic value of $g$ will depend on the microscopic model. Paper I has shown that in these cases we must have $\dprime{F}\neq 0$, and one may distinguish two subcases: if $\ddprime{F}=0$ then $g$ will be a constant (again, not necessarily unity), while if $\ddprime{F}\neq0$ the $g$ will have an implicit time dependence (except if there is fine-tuning of several model parameters).

As in Paper I, our approach is to look for asymptotic scaling solutions of power law form, defining
\begin{subequations}
\label{powerlaws}
\bq
    L_c &=& L_0\tau^\alpha \\
    v &=& v_0\tau^\beta \\
    J^2 &=& J_0^2 \tau^\gamma \\
    Q^2 &=& Q_0^2 \tau^\delta \\
    \xi_c &=& \xi_0 \tau^\varepsilon  \\
    W &=& W_0\tau^\zeta = \frac{\xi_0}{L_0}\tau^{\varepsilon-\alpha}\,,
\eq
\end{subequations}
where the physical solutions must be such that $\beta\leq 0$ and $\alpha\leq 1$. The scale factor is also assumed to have a power law form. In terms of conformal time, $\tau$,
\begin{equation}
    a = a_0 \tau^{\lambda}\,,\qquad \mathcal{H} = \frac{\dprime{a}}{a} = \frac{\lambda}{\tau}\,,
\end{equation}
or equivalently in terms of physical time, $t$,
\begin{equation}
    a(t) \propto t^{\frac{\lambda}{1+\lambda}}\,.
\end{equation}
This assumption is convenient numerically \cite{Simulations}, but also reasonable in the real universe, e.g. $\lambda=1$ corresponds to the radiation era and $\lambda=2$ to the matter era. The description in terms of conformal time and comoving length scales can be rephrased as a function of physical time and physical lengths. Any quantity that is scaling with respect to conformal time as a power law of exponent $\zeta$, will scale as a function of cosmic time as a power law with exponent $\zeta/(1+\lambda)$, while if a comoving length scales as $\tau^\kappa$ its physical counterpart will scale as $\tau^{\kappa+\lambda}$. In particular, this shows that linearly scaling comoving distances with respect to conformal time are equivalent to linearly scaling physical quantities with respect to cosmic time.

By assuming these power law solutions, the CVOS equations take the generic form
\begin{subequations}
\bq
    \alpha  &=&  \lambda \left[v_0^2\tau^{2\beta}-C_v\mathcal{K}\tau^{2\left(\alpha-\varepsilon\right)+\eta}\right] + \frac{g\tilde{c}}{2}\frac{v_0}{\xi_0} \tau^{1+\beta-\varepsilon} \\
    \beta &=& C_v\left[\frac{k_v}{v_0\xi_0}\tau^{1-\beta-\varepsilon}\left(1+2\mathcal{K}\tau^{2\left(\alpha-\varepsilon\right)+\eta}\right)\right]\nonumber\\
    && -C_v\left[2\lambda \left(1+\mathcal{K}\tau^{2\left(\alpha-\varepsilon\right)+\eta}\right)\right]\\
    \gamma &=& 2\left(\frac{v_0k_v}{\xi_0}\tau^{1+\beta-\varepsilon}-\lambda\right)\nonumber\\
    && -\rho\tilde{c}\frac{v_0}{\xi_0}\frac{\xi_0^2}{J_0^2L_0^2}\frac{1-g}{\dprime{F}-2Q_0^2\tau^\delta\ddprime{F}}\tau^{1+\beta+\varepsilon-2\alpha-\gamma}
    \\
    \delta  &=& 2\frac{\dprime{F}+2J_0^2 \tau^\gamma\ddprime{F}}{\dprime{F}+2Q_0^2 \tau^\delta\ddprime{F}}\left(\frac{v_0k_v}{\xi_0}\tau^{1+\beta-\varepsilon}-\lambda\right) 
    -\nonumber \\
    && -(1-\rho)\tilde{c}\frac{v_0}{\xi_0}\frac{\xi_0^2}{Q_0^2L_0^2}\frac{1-g}{\dprime{F}+2Q_0^2\tau^\delta\ddprime{F}}\tau^{1+\beta+\varepsilon-2\alpha-\delta}
    \\
    \varepsilon &=&\lambda v_0^2\tau^{2\beta}\left(1+\mathcal{K}\tau^{2\left(\alpha-\varepsilon\right)+\eta}\right)\nonumber\\
    && - \frac{v_0k_v}{\xi_0}\mathcal{K}\tau^{2\left(\alpha-\varepsilon\right)+\eta+1+\beta-\varepsilon}+\frac{\tilde{c}}{2}\frac{v_0}{\xi_0}\tau^{1+\beta-\varepsilon}
\eq
\end{subequations}
where for convenience we have defined
\begin{subequations}
\bq
    \mathcal{K} &=& \frac{L_0^2\mathcal{J}_0^2}{\xi_0^2}\dprime{F}\label{defnk}\\
    \mathcal{J}_0^2 &=&
    \begin{cases}
        Q_0^2+J_0^2&\text{, if }\delta=\gamma\\
        J_0^2&\text{, if }\delta<\gamma\\
        Q_0^2&\text{, if }\delta>\gamma \\
    \end{cases}\\
    \eta &=&
    \begin{cases}
    \gamma & \text{, if }\delta\leq\gamma\\
    \delta& \text{, if }\delta>\gamma\\
    \end{cases}\\
    C_v &=& 1-v^2\,.
\eq
\end{subequations}  
Assuming that the last of these is a constant is reasonable since for $\beta=0$, $1-v_0^2$ is a constant, while for $\beta<0$, $1-v_0^2t^{2\beta}\sim1$, which is also constant. Similarly the momentum parameter $k_v$ is velocity-dependent \cite{Martins_2002}, but it will also be a constant in both of these cases. Finally, in presenting the solutions below it has been assumed that all the pre-factor parameters in Eq.(\ref{powerlaws}) are non-zero. 

We also recall that Paper I noted the relevance of the factor $2(\alpha-\varepsilon)+\eta$, which appears associated with the first derivative $\dprime{F}$ and is related to Eq.(\ref{eq:lc_xic}), which we can write as
\begin{equation}
    \frac{\xi_0^2}{L_0^2}\tau^{2(\varepsilon-\alpha)}=F-2Q_0^2\tau^\delta\dprime{F}\,.
\end{equation}
Given that $F$ is not expected to vanish, this provides a constraint on the relation between the three exponents. For decaying or constant charge solutions ($\delta\leq0$) both length scales must evolve with a similar rate, $\alpha=\varepsilon$, as expected: the ratio of total and bare string energies will then be a constant (equal to unity in the Nambu-Goto limit). Conversely, growing charges are only possible if the overall characteristic length grows slower than the correlation length, again as it should be. Solutions where $\alpha>\varepsilon$, would require an unphysical $F=0$.

So far we have presented the general CVOS evolution equations under our general assumption of power law solutions for the various time-dependent quantities. In the following sections we sequentially present and discuss the specific asymptotic solutions, for the various relevant cases (e.g., without or with expansion), and following the same sequence of Paper I. For each such solution we also discuss the conditions under which each of these solutions can occur (these conditions can involve the cosmological expansion rate and model parameters). The goal is therefore to extend the classification of these solutions presented in Paper I, to include the case of biased solutions, It should be noted that the constraints found are strongly dependent on the case under the analysis. For instance, in some solutions the bias parameter $\rho$ must not be unity, while in other cases it is required that $\rho=1$. The only constraints that apply to all cases are those imposed by the causality requirement (meaning $\alpha\leq 1)$ and that the network velocity cannot exceed the speed of light (meaning $v<1$ and $\beta\leq0$).

%%%%%%%%%%%%%%%%%%%%%%%%%%%%%%%%%%%%%%%%%%%%%%%%%%%%%%%%%%%%%%%%%%%%%%%%%
\section{\label{03nomechanisms}Solutions without expansion}

We start with the simplest possible case, which corresponds to Minkowski spacetime. Here not only do we have $\lambda=0$ but the momentum parameter is also expected to vanish ($k_v=0$), reducing the equations to
\begin{subequations}
\bq
    \alpha  &=& \frac{g\tilde{c}}{2}\frac{v_0}{\xi_0}\tau^{1-\varepsilon}\\
    \beta  &=& 0\\
    \gamma &=& -\rho\tilde{c}\frac{v_0}{\xi_0}\frac{\xi_0^2}{J_0^2L_0^2}\frac{1-g}{\dprime{F}-2Q_0^2\tau^\delta\ddprime{F}}\tau^{1+\varepsilon-2\alpha-\gamma}
    \\
    \delta  &=& -(1-\rho)\tilde{c}\frac{v_0}{\xi_0}\frac{\xi_0^2}{Q_0^2L_0^2}\frac{1-g}{\dprime{F}+2Q_0^2\tau^\delta\ddprime{F}}\tau^{1+\varepsilon-2\alpha-\delta}
    \\
    \varepsilon &=& \frac{\tilde{c}}{2}\frac{v_0}{\xi_0}\tau^{1-\varepsilon} 
\eq
\end{subequations}
Note that we will assume that $g\neq1$, since the solutions with $g=1$ have already been discussed in Paper I. It is clear that the correlation length must be constant ($\varepsilon=1$) in all cases, implying that $\alpha=g<1$. This implies that in these solutions, energy losses from regions with charges and currents will be smaller than those from regions without them.

According to our previous discussion, the only additional physical solutions here are associated with growing charges, since $\delta=2(1-\alpha)=2(1-g)>0$. In these cases, one finds the solution where the current also grows:
\begin{subequations}
\bq
    L_c &=& L_0\tau^g\\
    \xi_c&=&\xi_0\tau \\
    J^2 &=& J_0^2\tau^{2-2g}\\
    Q^2 &=& Q_0^2\tau^{2-2g} \\
    v &=& v_0\,,
\eq
\end{subequations}
subject to the constraints
\begin{subequations}
\bq
    \tilde{c}&=&\frac{2\xi_0}{v_0}\\
    \frac{1-\rho}{\rho} &=& \frac{J_0^2}{Q_0^2}\\
    \rho&\neq& 1\\
    \ddprime{F}&=&0\,.
\eq
\end{subequations}
It is worthy of note that the ratio of the charge and current is determined by the bias parameter $\rho$, as one might have expected. If $\rho=1/2$ we have a chiral solution, but non-chiral solutions, with any finite value of the charge-to-current ratio other than unity are also allowed in principle. However, the case $\rho=1$, where the charge would be constant while the current would grow, is not a solution.

On the other hand, in the opposite limit $\rho=0$, there is a different solution, where the charge still grows, but the current is kept constant, 
\begin{subequations}
\bq
    L_c &=& L_0\tau^g\\
    \xi_c&=&\xi_0\tau \\
    J^2 &=& J_0^2\\
    Q^2 &=& Q_0^2\tau^{2-g} \\
    v &=& v_0\,,
\eq
\end{subequations}
subject to the constraints
\begin{subequations}
\bq
    \tilde{c}&=&\frac{2\xi_0}{v_0}\\
    \rho &=& 0\\
    \ddprime{F}&=&0\,.
\eq
\end{subequations}

Physically, the interpretation of these solutions is clear: $g<1$ means that long string regions containing charges or currents are less likely to be lost into loops than those without them. This charge, and for $\rho\neq0$ also the current, therefore accumulates in the network, while the decreased energy losses means that $L_c$ grows more slowly than $\xi_c$, so the solution differs from the usual linear scaling. While simulating these networks in Minkowski space is numerically more challenging than in expanding universe, due to the enhanced amount of radiation in the simulation box, such simulations could easily provide a convenient way to measure the bias parameter $g$ in specific field theory models.

Still, we note that taking the limit $g\to1$ in either of these solutions recovers the solution given by Eq. (21) of Paper I.

%%%%%%%%%%%%%%%%%%%%%%%%%%%%%%%%%%%%%%%%%%%%%%%%%%%%%%%%%%%%%%%%%%%%%%%%%
\section{\label{03nolosses}Cosmological solutions without energy loss mechanisms}

Paper I next considered the case where cosmological expansion is included, but there are no other energy loss mechanisms, in other words we have $\tilde{c}=0$. Although we keep an analogous section here to maintain the structures of the two papers synchronized, it is clear that there are no additional solutions to report for arbitrarily biased networks, beyond those discussed therein. Note that strictly speaking this does not mean that no further solutions are possible, but only that they are asymptotically equivalent to the ones already discussed for the unbiased networks.

Nevertheless, it is instructive to very briefly mention here the most salient outcomes of the analysis of the solutions on the corresponding section in Paper I, since they are relevant for the results presented in the following section. A first point is that the only allowed solutions have $\varepsilon=1+\beta\leq1$. This includes Nambu-Goto type linear scaling (where charge and current both disappear), highlighting the amply established fact that linear scaling is possible even in the absence of loop production, provided the expansion rate is fast enough \cite{Martins_2016,Almeida_2021}. A second point is that there are different branches of solutions depending on whether $\dprime{F}=0$ or $\dprime{F}\neq0$, highlighting the fact that different microphysical models can lead to different cosmological evolutions.

Finally, the third point was already mentioned in the introduction: broadly speaking (with caveats related to conditions applicable to each solution branch), scaling solutions can be divided into three classes, primarily determined by the cosmological expansion rate. Fast expansion rate solutions asymptote to the Nambu-Goto case (with charge and current disappearing), while slow expansion rates can allow the charge and/or current to dominate the network dynamics and prevent linear scaling. Under the separation of variables implicit in the analytical model under study, we have shown that the latter scenario is also only found for a specific expansion rate. In between these two scaling classes there is a third one, valid for a single value of the expansion rate, in which the network is in full scaling, with a constant charge and current adding to the standard linear scaling behaviour. In Paper I we have shown that in the unbiased case the critical expansion rates correspond to $\lambda=2/3$ and to the matter-dominated era ($\lambda=2$) for the cases of decaying and constant velocities respectively. We will revisit this point in the following section.

%%%%%%%%%%%%%%%%%%%%%%%%%%%%%%%%%%%%%%%%%%%%%%%%%%%%%%%%%%%%%%%%%%%%%%%%%
\section{\label{04losses}Biased solutions with energy loss mechanisms}

Finally, we discuss the possible solutions when energy loss mechanisms are allowed, by setting $\tilde{c}\neq0$. Following the earlier discussion, we also note that the solutions for $g=1$ discussed in Paper I should also be interpreted as applying to any networks with biased charge loss mechanisms in which the charge and current decay away. Conversely, setting $g\neq1$ has the consequence of requiring solutions with non-decaying charges and/or currents. 

Although it may not be apparent at first sight, biased charge loss mechanisms effectively prevent solutions which extend preciously discussed ones with constant velocity and $\dprime{F}=0$. This can be seen by noting that, were this to be true, then one would have from the characteristic length equations
\bsq
\bq
    & \alpha = \lambda v_0^2\tau^{2\beta} + g\frac{\tilde{c}}{2}\frac{v_0}{\xi_0}  \\
    & \varepsilon = \lambda v_0^2\tau^{2\beta}+\frac{\tilde{c}}{2}\frac{v_0}{\xi_0}\,.
\eq
\esq
For constant values of $g$ (but still considering $g\neq1$), the expressions above imply $\alpha\neq\varepsilon$, while the consistency relation would also yield $L_cW=L_c\sqrt{F-2Q^2\dprime{F}}=L_c\sqrt{F}=\xi_c$ and $\alpha=\varepsilon$. This is clearly incompatible and hence all scaling solutions must be such that $\dprime{F}\neq 0$. This provides a phenomenological link between the microphysics of specific models and their macroscopic dynamics. Further exploring the equations reveals that all allowed solutions in this class must have $\varepsilon=1+\beta$, as was the case in the solutions discussed in the corresponding section of Paper I. 

We start with the subset of solutions with decaying velocities, $\beta<0$. Here there is one solution that is absent for unbiased networks, where both charge and current are allowed to grow
\bsq
\label{solutionC1}
\bq
    L_c &=& L_0\tau^{\frac{\lambda}{2}+ \frac{g\tilde{c}}{2}\frac{v_0}{\xi_0}}\\
    \xi_c&=&\xi_0\tau^{1-\lambda} \\
    J^2 &=& J_0^2\tau^{2-3\lambda-g\tilde{c}\frac{v_0}{\xi_0}}\\
    Q^2 &=& Q_0^2\tau^{2-3\lambda-g\tilde{c}\frac{v_0}{\xi_0}} \\
    v &=& v_0\tau^{-\lambda}\,,
\eq
\esq
subject to the constraints
\bsq
\label{constraintC1}
\bq
    \lambda &<& \lambda_c\equiv\frac{2}{3+g\tilde{c}/k_v}\\
    \mathcal{K} &=& -\frac{1}{2}\\
    \frac{v_0}{\xi_0} &=& \frac{2-\lambda}{2k_v+g\tilde{c}}\\
    \frac{\rho}{1-\rho} &=& \frac{J_0^2}{Q_0^2} \,,
\eq
\esq
with the last of these conditions only applying if one has $\ddprime{F}=0$. Note that $\mathcal{K}\propto F'$, cf. Eq. (\ref{defnk}), explicitly showing that the latter can't vanish.

This solution is only possible when the expansion rate is sufficiently low, and one can easily see that by making the replacements $g\to1$ or $\tilde{c}\to0$ one recovers the critical expansion rate identified in Paper I. Again, the ratio of the charge and current is determined by the bias parameter $\rho$, and both chiral and non-chiral solutions are allowed.

The upper bound for the expansion rate, $\lambda_c$, is related to the previously highlighted value of $\lambda=2/3$, and in fact a constant charge and current solution that generalises a solution from Paper I is given by
\bsq
\label{solutionC2}
\bq
    L_c &=& L_0\tau^{1-\lambda}\\
    \xi_c&=&\xi_0\tau^{1-\lambda}\\
    J^2 &=& J_0^2\\
    Q^2 &=& Q_0^2 \\
    v &=& v_0\tau^{-\lambda}\,,
\eq
\esq
subject to the constraints
\bsq
\label{constraintC2}
\bq
    \lambda &=& \frac{2+2\left(1-g\right)c/k_v}{3+\left(3-2g\right)c/k_v}\\
    \mathcal{K} &=& -\frac{1}{2}\\
    \rho &\neq& 0\,,
\eq
\esq
which manifestly reduces to solutions discussed in Paper I when $g=1$. In the general case under present consideration, there is an additional constraint given by
\begin{equation}
\begin{split}
    & \rho\left(\frac{\dprime{F}\left(Q_0^2+J_0^2\right)+2Q_0^2\ddprime{F}\left(Q_0^2-J_0^2\right)}{J_0^2(\dprime{F}-2Q_0^2\ddprime{F})} \right)  =\\
    & 1-4\left(Q_0^2-J_0^2\right) \frac{Q_0^2L_0^2}{\xi_0^2}\ddprime{F}\,.
\end{split}  
\end{equation}
Although cumbersome, this constraint is particularly interesting since in the chiral limit where $Q_0^2=J_0^2$ it reduces to
\begin{equation}
    \rho=\frac{\dprime{F}-2Q_0^2\ddprime{F}}{2\dprime{F}}
\end{equation}
which can easily be seen to be compatible with the chiral limit if one further imposes $\ddprime{F}=0$, yielding $\rho=1/2$.

\begin{figure*}
    \centering
    \includegraphics[width=0.32\textwidth]{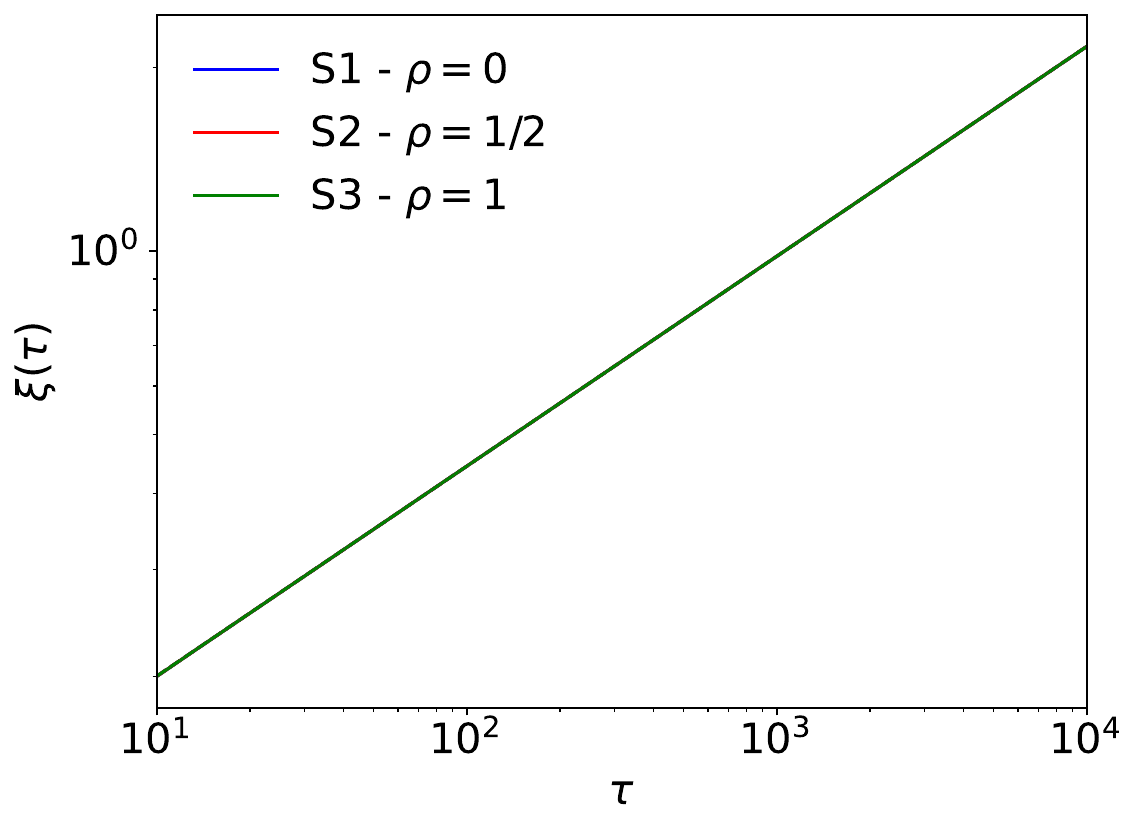}
    \includegraphics[width=0.32\textwidth]{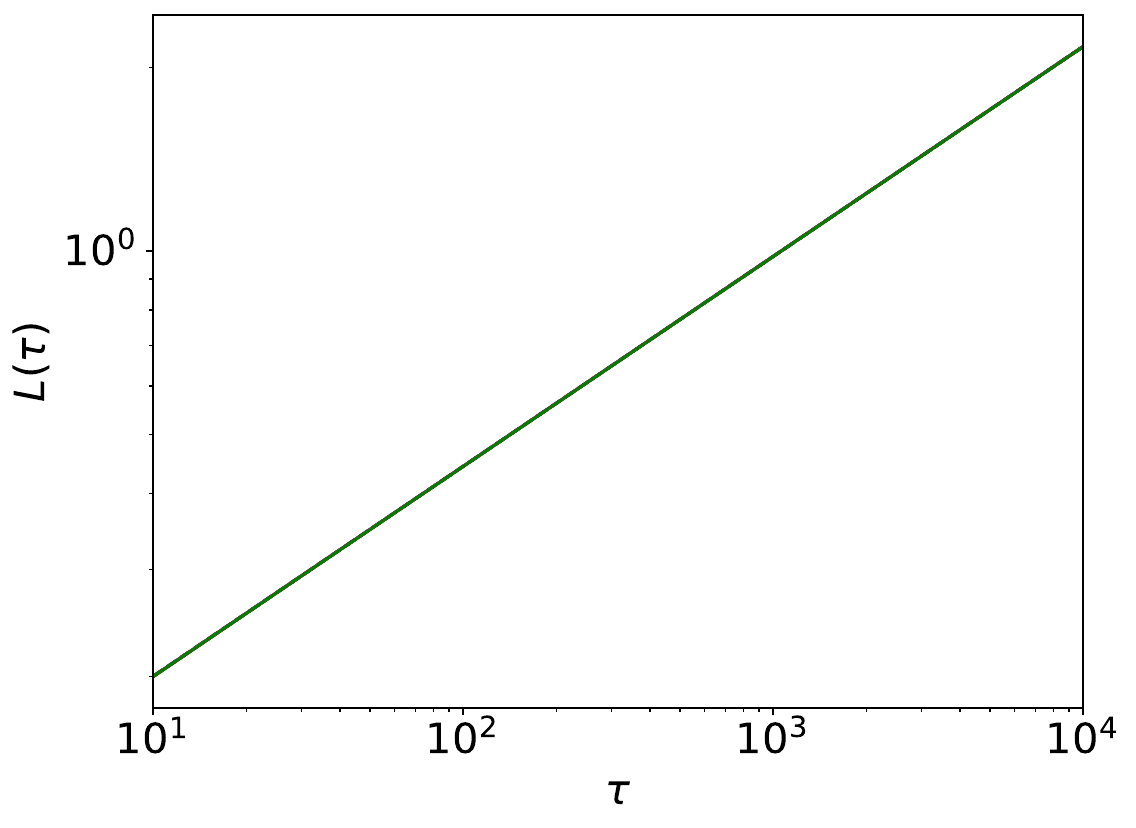}
    \includegraphics[width=0.32\textwidth]{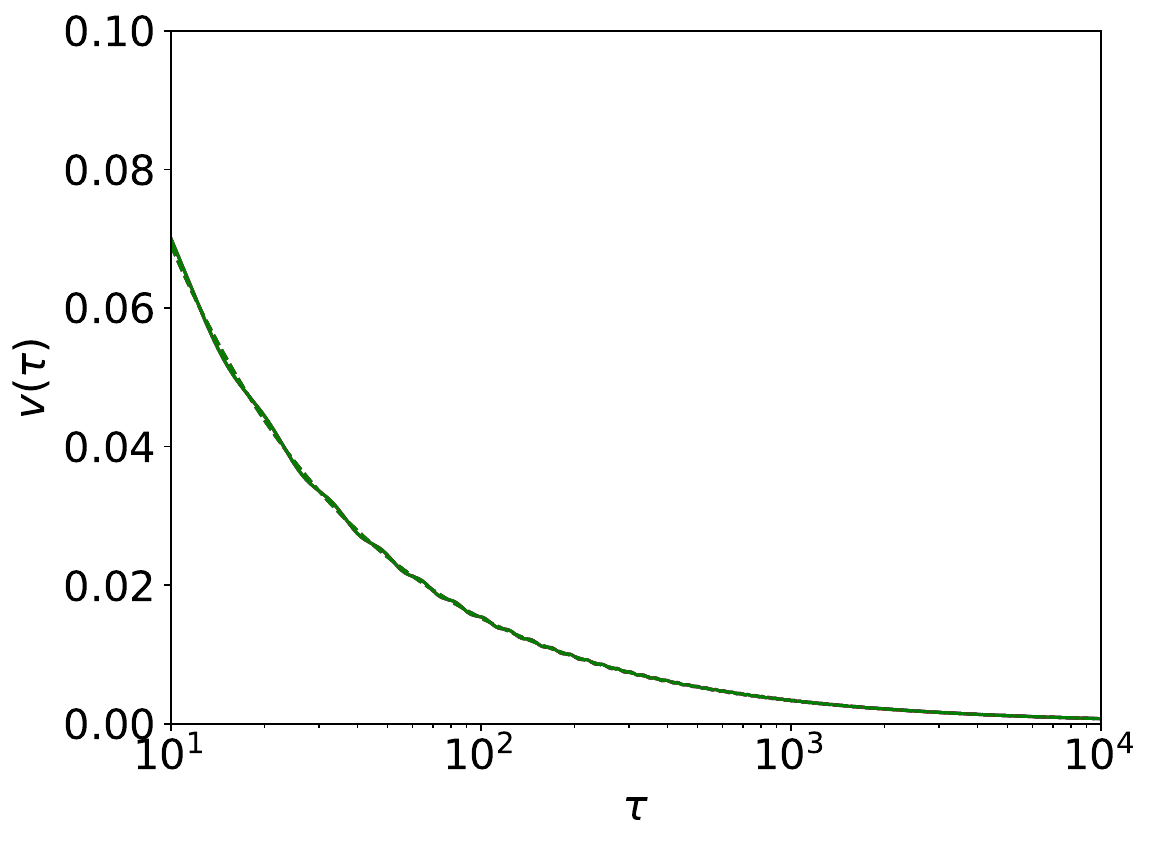}
    \includegraphics[width=0.32\textwidth]{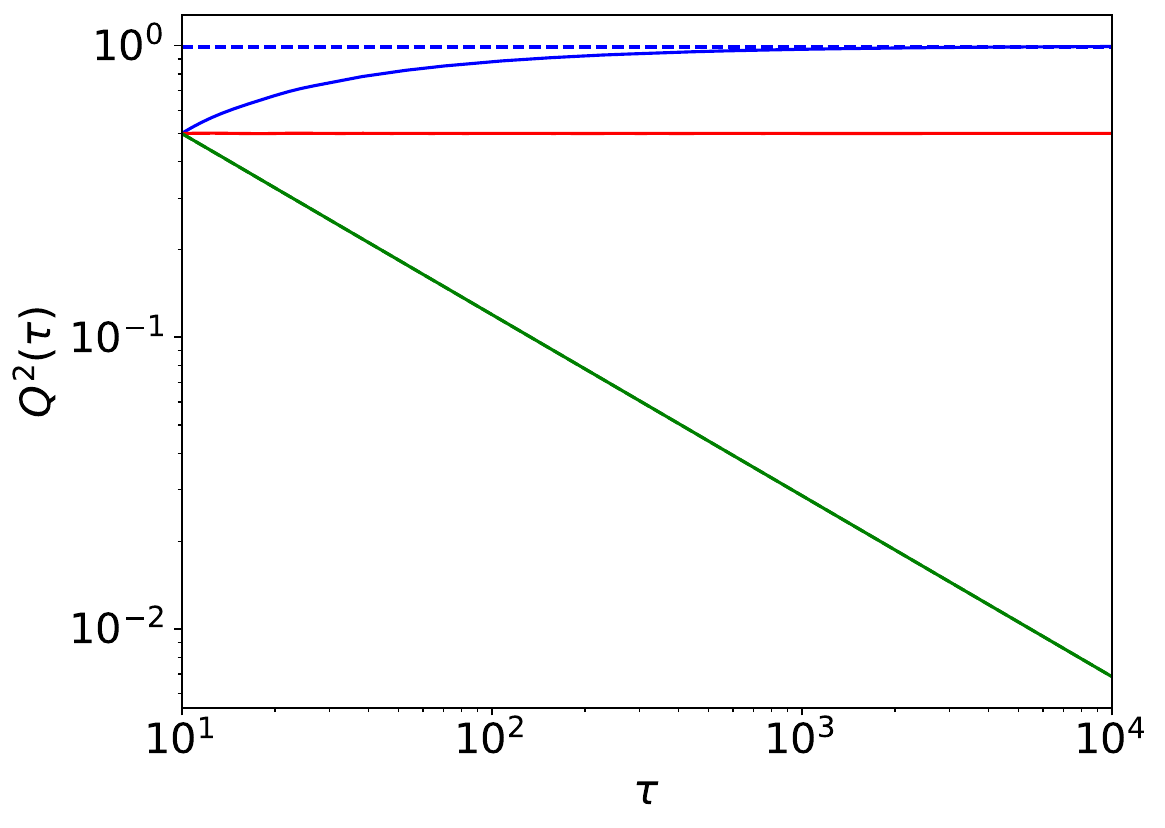}
    \includegraphics[width=0.32\textwidth]{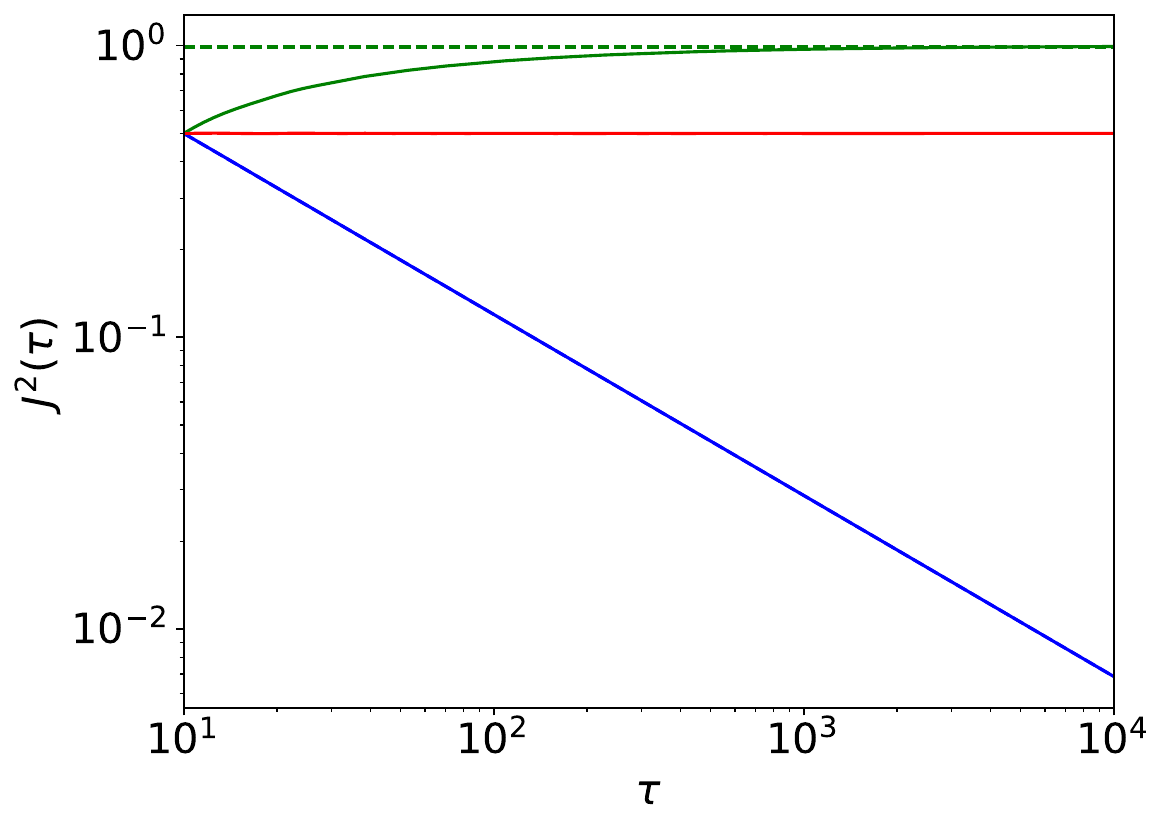}
    \includegraphics[width=0.32\textwidth]{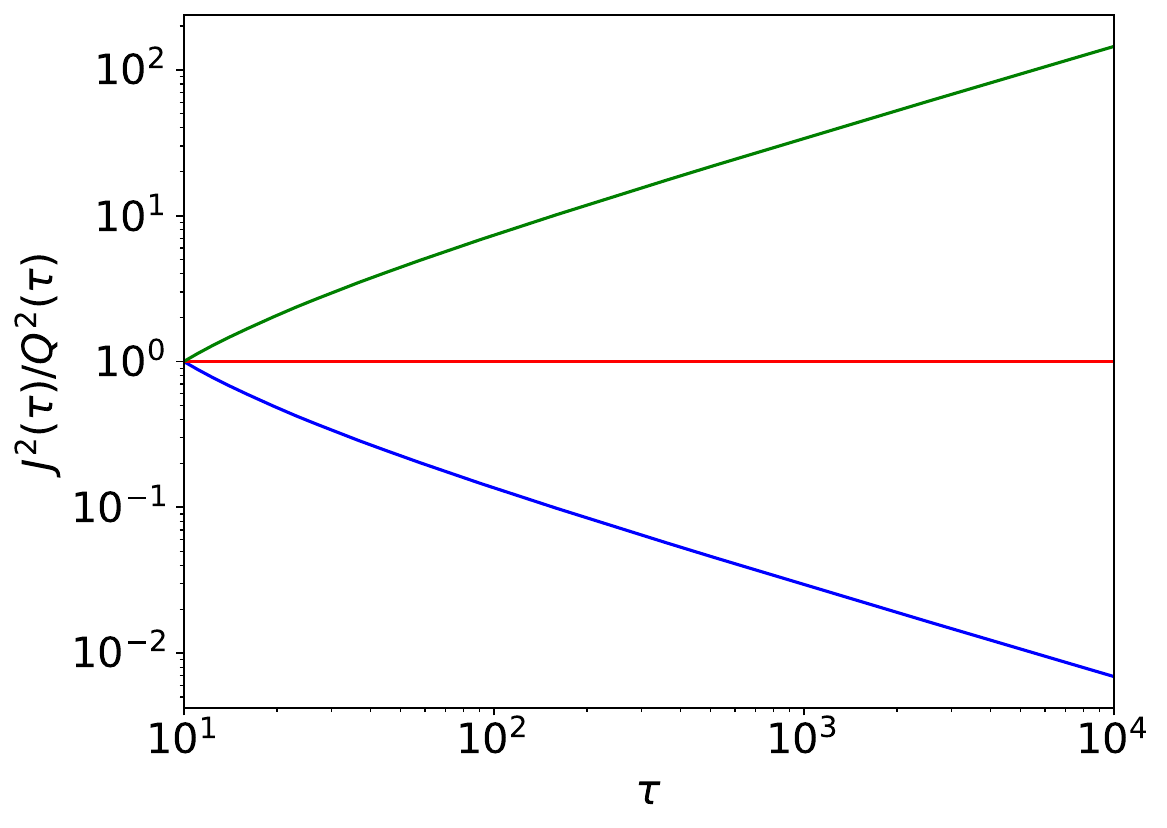}
    \caption{Network evolution as obtained by numerically solving the CVOS equations for $\tilde{c}=g=k_v=0.1$ and the expansion rate defined by Eq. \ref{constraintC2}. The three cases correspond to the same initial conditions, but with different values of the charge and current bias parameter $\rho$: (Blue) $\rho=0$; (Red) $\rho=1/2$; (Green) $\rho=1$.}
    \label{fig01}
\end{figure*}

A word of caution should be placed here, though, since this solution allows biased networks but not maximally biased ones, meaning those with $\rho=0$ or $\rho=1$. In these two extreme cases, different solutions apply, differing only on the charge and current behaviour. For the case $\rho=0$ the solution is
\bsq
\label{solutionC3}
\bq
    J^2 &=& J_0^2\tau^{-\frac{2(1-g)\tilde{c}}{k_v+(1-g)\tilde{c}}\lambda}=J_0^2\tau^{-\frac{4(1-g)\tilde{c}}{3k_v+(3-2g)\tilde{c}}}\\
    Q^2 &=& Q_0^2 \,,
\eq
\esq
and in the opposite limit $\rho=1$ we have
\bsq
\label{solutionC4}
\bq
    J^2 &=& J_0^2\\
    Q^2 &=& Q_0^2\tau^{-\frac{2(1-g)\tilde{c}}{k_v+(1-g)\tilde{c}}\lambda} =  Q_0^2\tau^{-\frac{4(1-g)\tilde{c}}{3k_v+(3-2g)\tilde{c}}}\,.
\eq
\esq
Both cases are subject to the additional constraint (beyond the one on $\rho$ itself)
\begin{equation}
    \label{constraintC4}
    \frac{2(1-g)\tilde{c}/k_v}{1+(1-g)\tilde{c}/k_v}>0\,.
\end{equation}
Here, the charge (or current) is still preserved, just like in Eqs. \ref{solutionC2}, but now the current (or charge) decays away. This effect can seen in Figure \ref{fig01}, where the same initial conditions evolve towards different asymptotic solutions depending one the value of $\rho$.

%%%%%%%%%%%%%%%%%%%%%%%%%%%%%%%%%%%%%%%%%%%%%%%%%%%%%%%%%%%%%%%%%%%
%%%%%%%%%%%%%%%%%%%%%%%%%%%%%%%%%%%%%%%%%%%%%%%%%%%%%%%%%%%%%%%%%%%

Next we focus on the branch of scaling solutions where the network velocity is constant. Here it is useful to distinguish three sub-branches, where charge and current evolve in a similar way ($\gamma=\delta$), where charge dominates ($\delta>\gamma$) and where current dominates ($\gamma>\delta$). In any case, all solutions are such that $\eta=2(\varepsilon-\alpha)=2(1-\alpha)$.

In the first sub-branch, there is a constant charge and current solution, where the characteristic lengths scale linearly with conformal time
\bsq
\label{solutionC6}
\bq
    L_c &=& L_0\tau\\
    \xi_c&=&\xi_0\tau\\
    J^2 &=& J_0^2\\
    Q^2 &=& Q_0^2 \\
    v &=& v_0\,,
\eq
\esq
subject to the constraints
\bsq
\label{constraintC6}
\bq
    \lambda &=& \frac{2-\tilde{c}/k_v\left(1-g\right)/\mathcal{K}}{1+\tilde{c}/k_v}\\
    \rho &=&
    \frac{J_0^2}{Q_0^2+J_0^2}\frac{\dprime{F}-2Q_0^2\ddprime{F}}{\dprime{F}}\\
    2\frac{\ddprime{F}}{\dprime{F}}\frac{Q_0^2-J_0^2}{Q_0^2+J_0^2} +\frac{1}{Q_0^2}
    &=&
    \rho\left(\frac{1}{Q_0^2}+\frac{1}{J_0^2}\frac{\dprime{F}+2Q_0^2\ddprime{F}}{\dprime{F}-2Q_0^2\ddprime{F}}\right)\,.
\eq
\esq
This is therefore a full scaling solution, which only holds for a single expansion rate. For $g=1$ the expansion rate from Paper I is recovered. If $\ddprime{F}=0$ the value $\rho=1/2$ again corresponds to chirality. With negligible energy losses the critical expansion rate would be the matter era ($\lambda=2$) as expected, but energy losses and biases will impact its value. For this critical value to correspond to the radiation era one needs
\be
\tilde{c}=\frac{k_v}{1+(1-g)\mathcal{K}}\,.
\ee

There are also two solutions where both current and charge grow over time
\bsq
\label{solutionC7}
\bq
    L_c &=& L_0\tau^\alpha\\
    \xi_c&=&\xi_0\tau\\
    J^2 &=& J_0^2\tau^{2-2\alpha}\\
    Q^2 &=& Q_0^2\tau^{2-2\alpha} \\
    v &=& v_0\,,
\eq
\esq
subject to the constraints
\bsq
\bq
    \alpha &=& \lambda \left(v_0^2-C\mathcal{K}\right) + \frac{g\tilde{c}}{2}\frac{v_0}{\xi_0}\\
    \frac{v_0k_v}{\xi_0} &=& 2\lambda v_0^2\frac{1+\mathcal{K}}{1+2\mathcal{K}} =
    \frac{2}{1+\tilde{c}/k_v}\,.
\eq
\esq
that have different additional constraints depending on whether $\ddprime{F}=0$ or $\ddprime{F}\neq 0$. For $\ddprime{F}=0$, it must be that
\bsq
\label{constraintC7}
\bq
    \lambda &<& \frac{1}{v_0^2-\mathcal{C}_v\mathcal{K}}\frac{1+\left(1-g\right)\tilde{c}/k_v}{1+\tilde{c}/k_v}\\
    \frac{Q_0^2}{J_0^2}&=&\frac{1-\rho}{\rho}\,.
\eq
\esq
while for $\ddprime{F}\neq 0$ it must be that
\bsq
\label{constraintC8}
\bq
    \lambda &<& \frac{2}{1+\tilde{c}/k_v}\\
    Q_0^2&=&J_0^2\,.
\eq
\esq

The sub-branch where current eventually dominates the evolution is simpler: it requires $\rho=1$ and a single solution is possible
\bsq
\label{solutionC9}
\bq
    L_c &=& L_0\tau\\
    \xi_c&=&\xi_0\tau\\
    J^2 &=& J_0^2\\
    Q^2 &=& Q_0^2 \tau^\delta\\
    v &=& v_0\,,
\eq
\esq
and is subject to the constraints
\bsq
\label{constraintC9}
\bq
    \lambda &=& \frac{2-\tilde{c}/k_v\left(1-g\right)/\mathcal{K}}{1+\tilde{c}/k_v}\\
    \delta &=& \frac{\tilde{c}}{k_v}\frac{\dprime{F}+2J_0^2 \ddprime{F}}{\dprime{F}}\frac{1-g}{\mathcal{K}}\frac{2}{1+\tilde{c}/k_v}<0\\
    \delta &=& -2\mathcal{C}\frac{\dprime{F}+2J_0^2 \ddprime{F}}{\dprime{F}}\frac{1+\mathcal{K}}{1+2\mathcal{K}}\lambda<0\\
    \rho&=&1\,.
\eq
\esq
In this case the current is constant, the charge decays, and the characteristic lengths scales linearly. These restrictions remove the role of $\ddprime{F}$ in further distinguishing between different solutions. 

Conversely, the last sub-branch requires $\rho=0$ and has charge dominating over current. In this case, the value of $\ddprime{F}$ is relevant once more, since it effectively decouples different solutions. If $\ddprime{F}$ does not vanish, one finds the solution
\bsq
\label{solutionC10}
\bq
    L_c &=& L_0\tau\\
    \xi_c&=&\xi_0\tau\\
    J^2 &=& J_0^2\tau^\gamma\\
    Q^2 &=& Q_0^2\\
    v &=& v_0\,,
\eq
\esq
subject to the constraints
\bsq
\label{constraintC10}
\bq
    \lambda &=& \frac{2-\tilde{c}/k_v\left(1-g\right)/\mathcal{K}}{1+\tilde{c}/k_v}\\
    \gamma &=& 2\left(\frac{v_0k_v}{\xi_0}-\lambda\right)<0\\
    \rho&=&0\,,
\eq
\esq
while for a vanishing $\ddprime{F}=0$ the possible solutions is given by
\bsq
\label{solutionC11}
\bq
    L_c &=& L_0\tau^\alpha\\
    \xi_c&=&\xi_0\tau\\
    J^2 &=& J_0^2\tau^\gamma\\
    Q^2 &=& Q_0^2\tau^\delta\\
    v &=& v_0\,,
\eq
\esq
subject to the constraints
\bsq
\label{constraintC11}
\bq
    \lambda &=& \frac{v_0}{1-v_0^2}\frac{1}{\xi_0\left(1+\mathcal{K}\right)}\left(k_v-\frac{\tilde{c}}{2}\frac{1-g\left(1+\mathcal{K}\right)}{\mathcal{K}}\right)\\
    0&>&\tilde{c}\frac{1-g}{\mathcal{K}}\\
    \gamma  &=&  2\left(\frac{v_0k_v}{\xi_0}-\lambda\right)\\
    \delta  &=& \gamma
    -\frac{\tilde{c}v_0}{\xi_0}\frac{1-g}{\mathcal{K}}\\
    \alpha &=& \lambda \left(v_0^2-C\mathcal{K}\right) + \frac{g\tilde{c}}{2}\frac{v_0}{\xi_0}\\
    \delta &=& 2-2\alpha>0\\
    \rho&=&0\,.
\eq
\esq

\begin{figure*}
    \centering
    \includegraphics[width=0.32\textwidth]{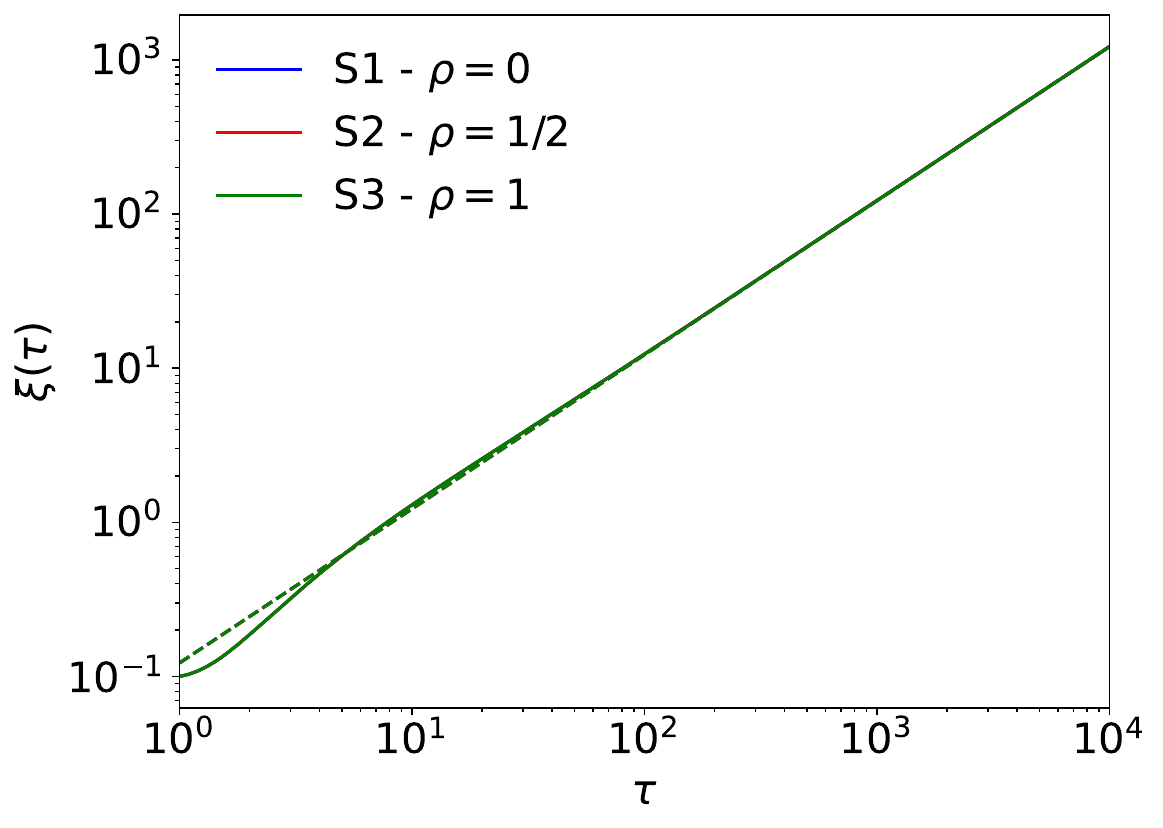}
    \includegraphics[width=0.32\textwidth]{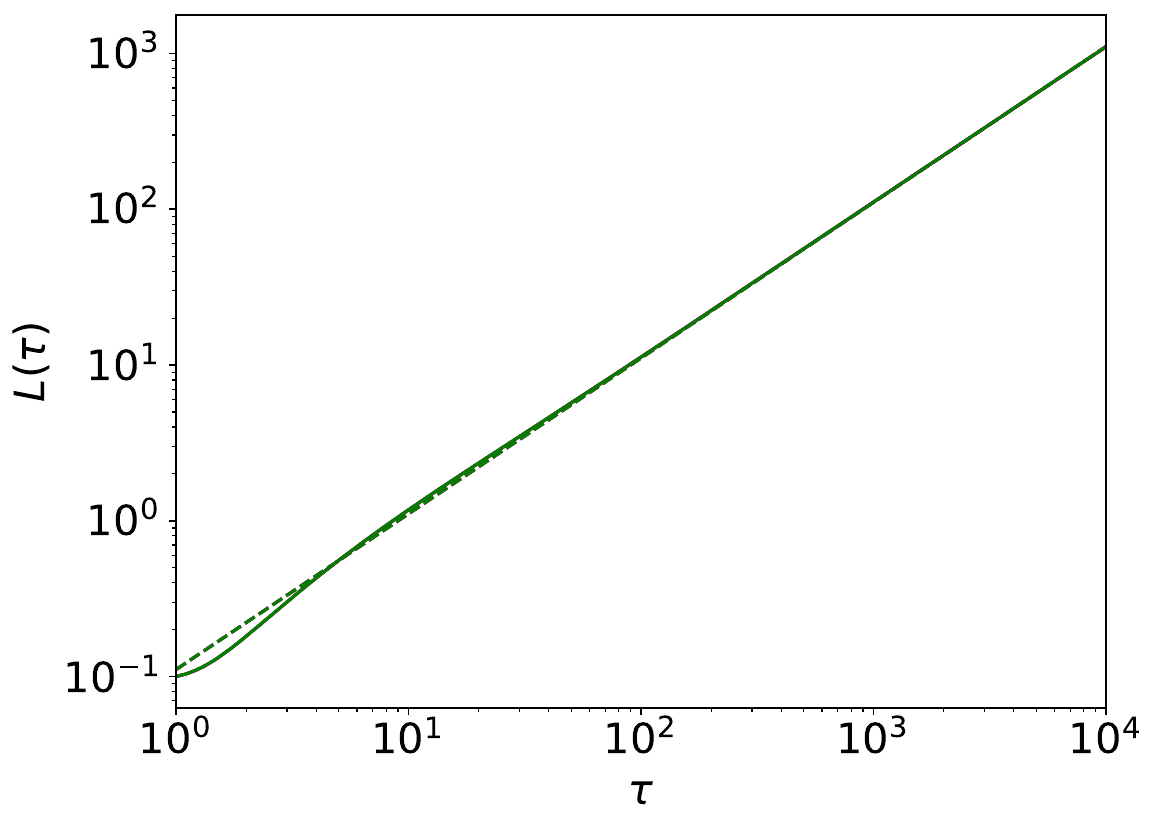}
    \includegraphics[width=0.32\textwidth]{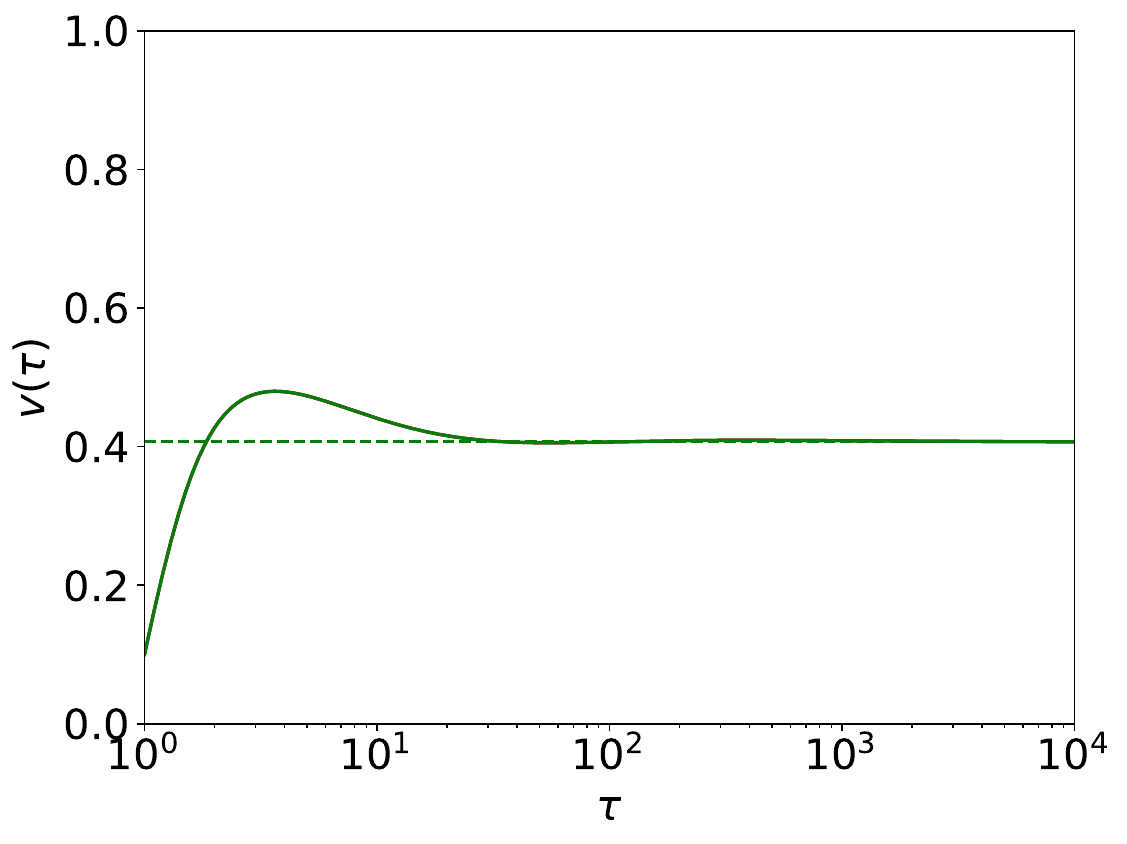}
    \includegraphics[width=0.32\textwidth]{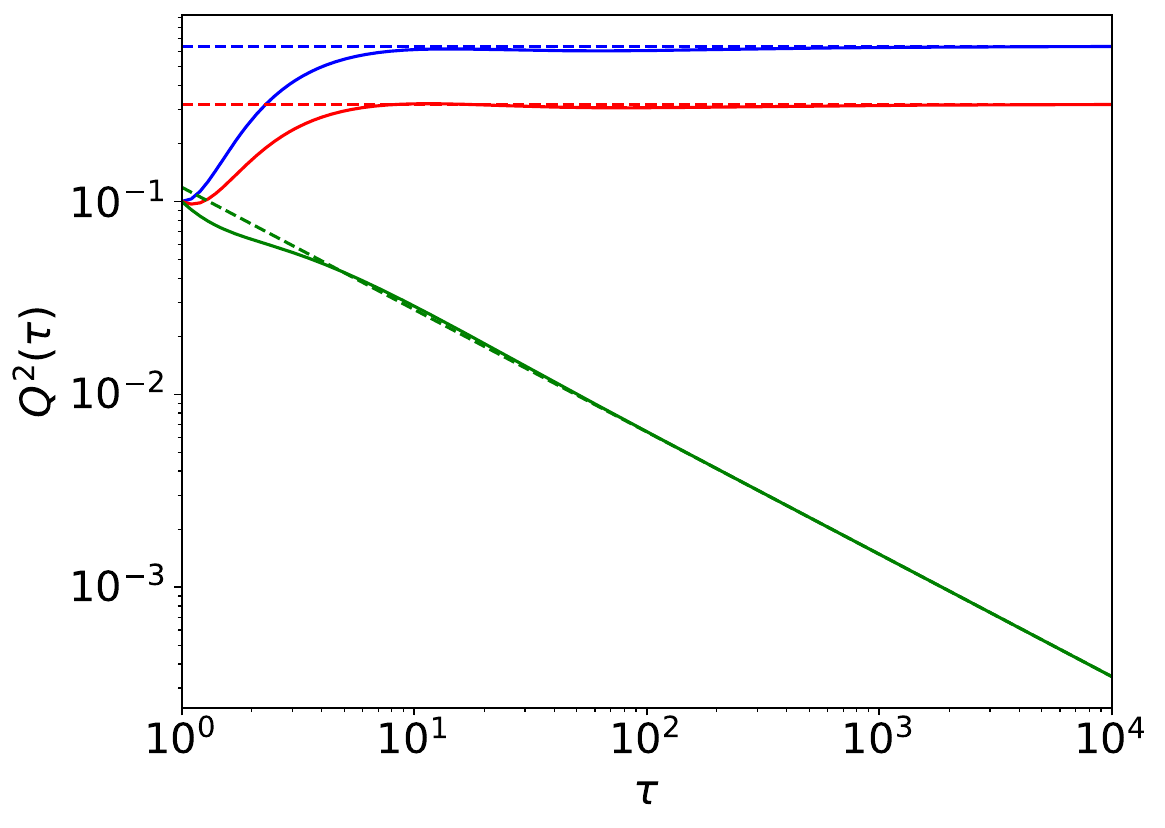}
    \includegraphics[width=0.32\textwidth]{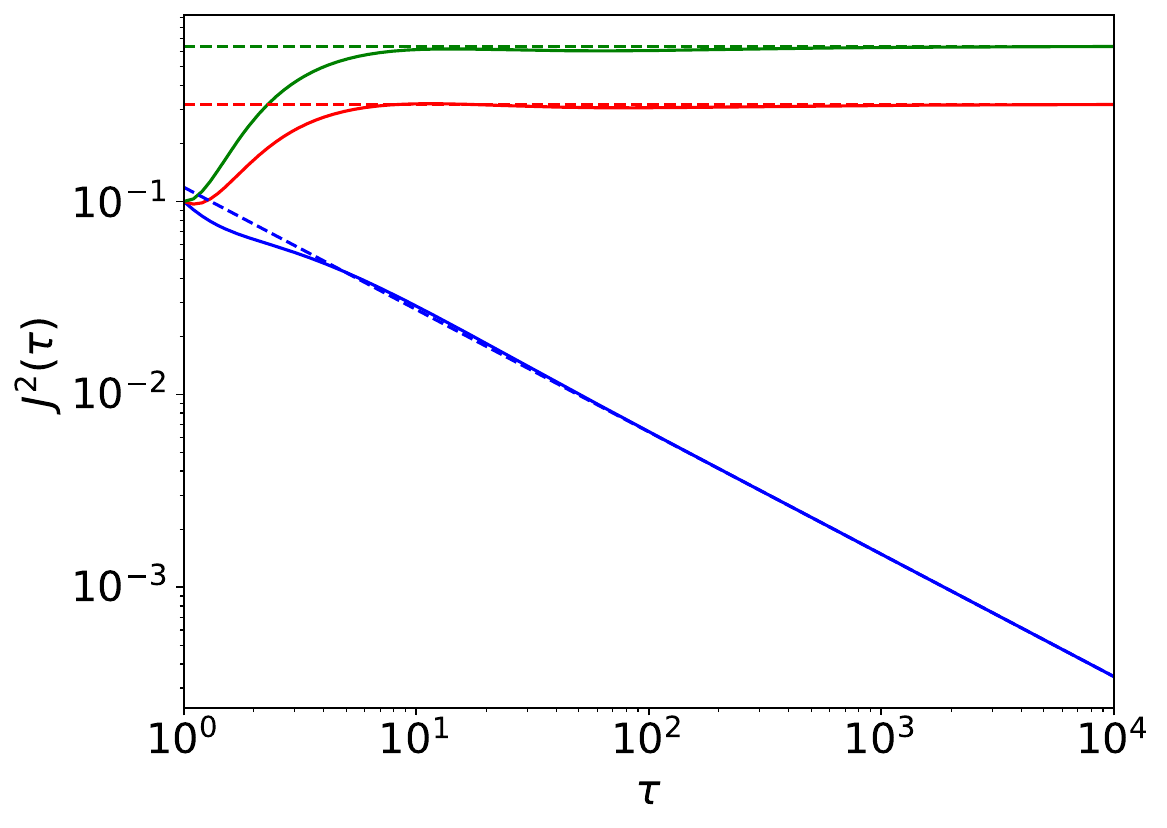}
    \includegraphics[width=0.32\textwidth]{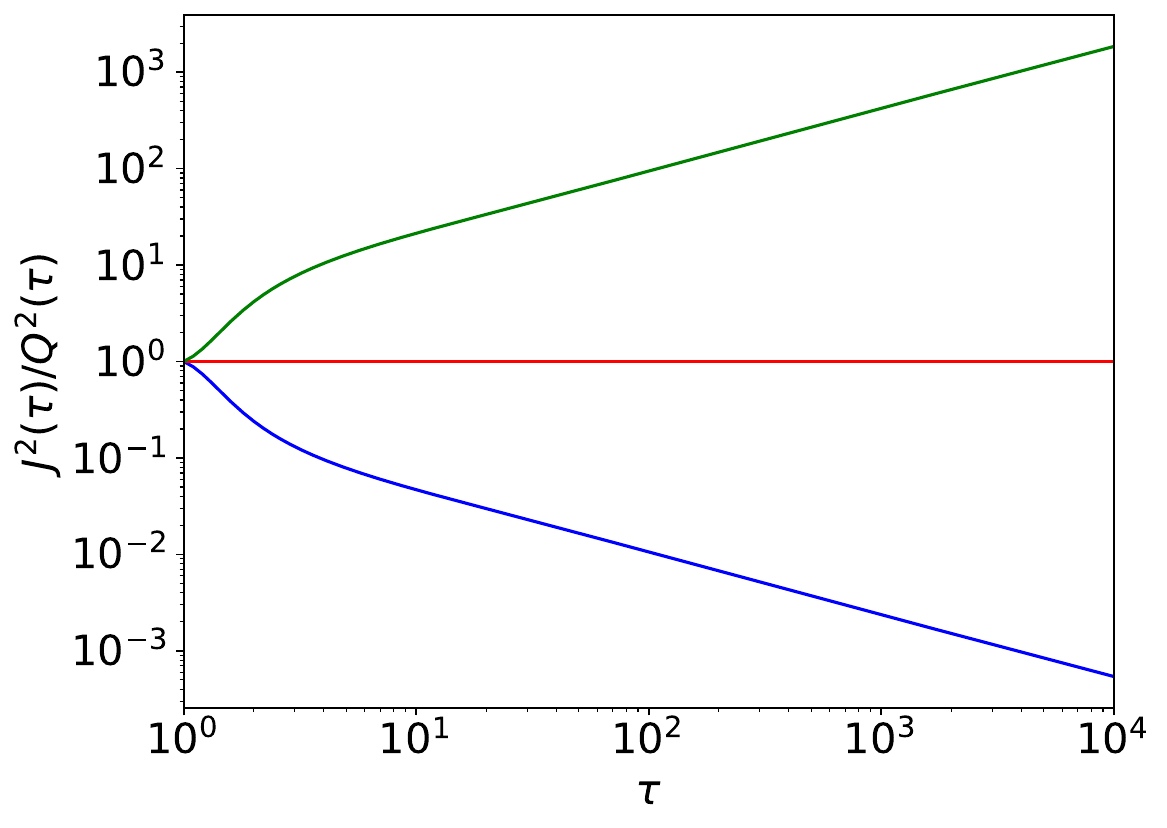}
    \caption{Numerical evolution of networks analogous to those of Figure \ref{fig01} but for the constant velocity and $F^\prime=-1/2\neq0$ branch, with $\tilde{c}=0.5$ and $g=0.9$.  The three cases correspond to the same initial conditions, but with different charge and current bias parameter $\rho$: (Blue) $\rho=0$; (Red) $\rho=1/2$; (Green) $\rho=1$.}
    \label{fig02}
\end{figure*}

Both of these cases, having $\rho=0$, correspond to the opposite maximally biased limit to that of Eqs. \ref{solutionC9} (for which $\rho=1$). If $\ddprime{F}\neq0$, then charge is not allowed to grow and must remain constant, while currents must decay and the network's characteristic lengths scale linearly, as given by Eqs. \ref{solutionC10}. On the other hand, if $\ddprime{F}=0$, a more general solution is possible, as given by Eqs. \ref{solutionC11}. A numerical comparison of the three different sub-branches is presented in Figure \ref{fig02}.

\begin{figure*}
    \centering
    \includegraphics[width=0.32\textwidth]{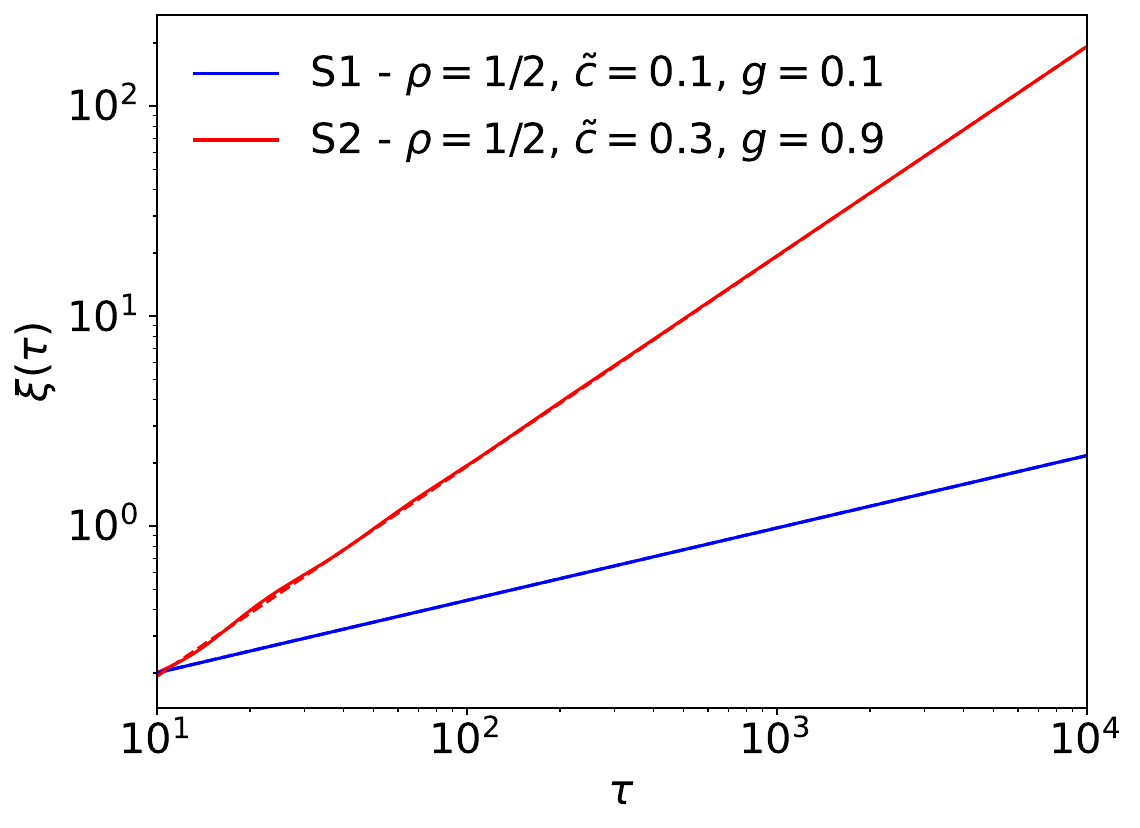}
    \includegraphics[width=0.32\textwidth]{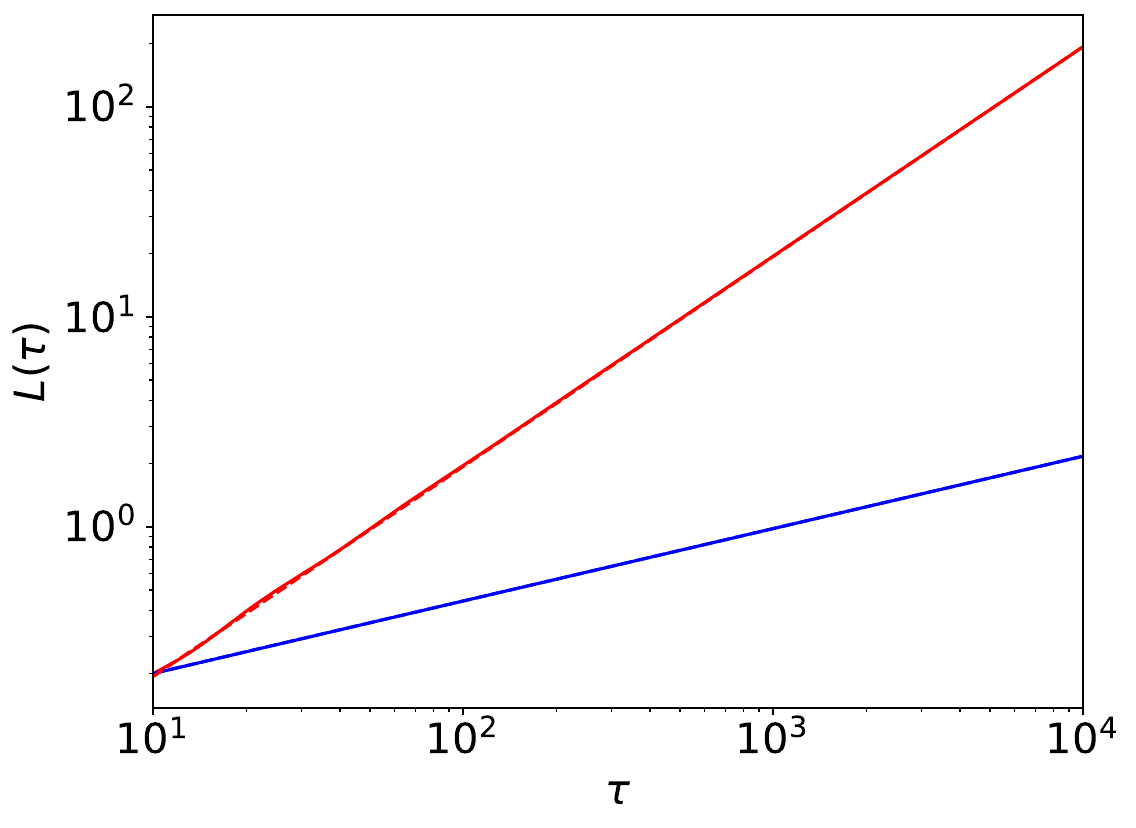}
    \includegraphics[width=0.32\textwidth]{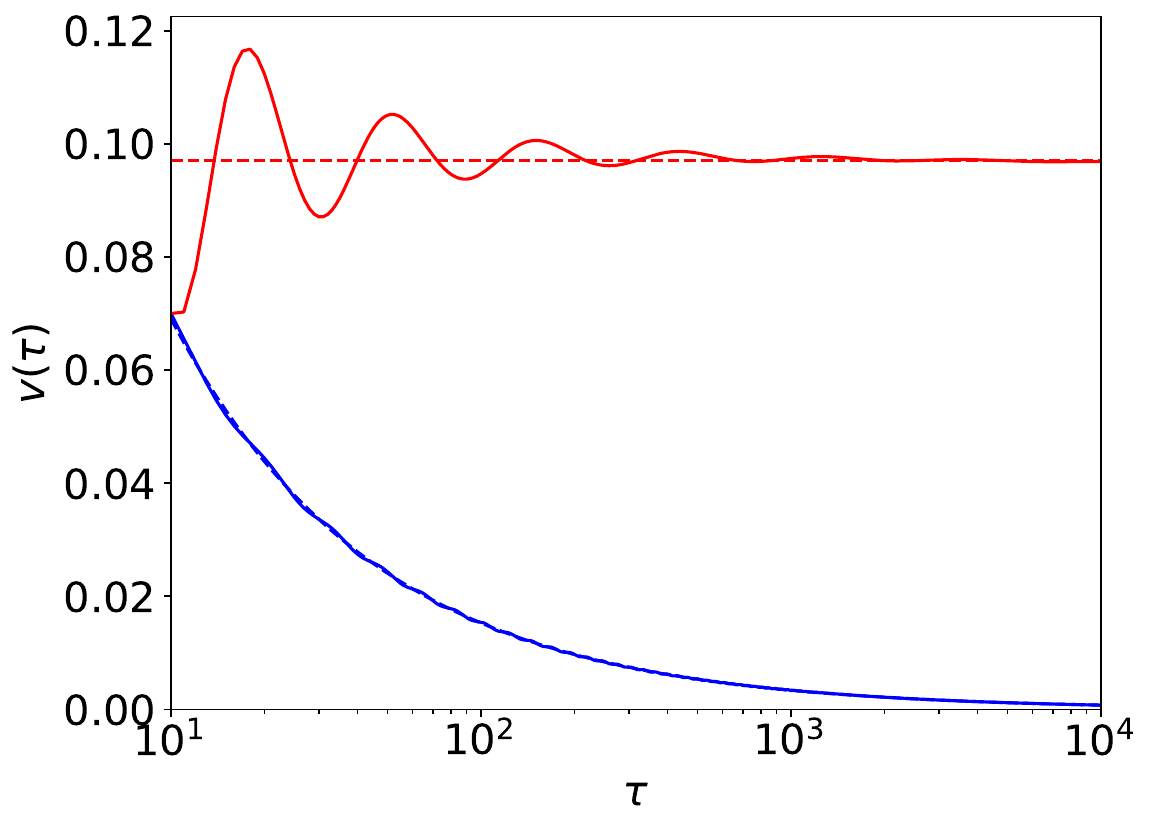}
    \includegraphics[width=0.32\textwidth]{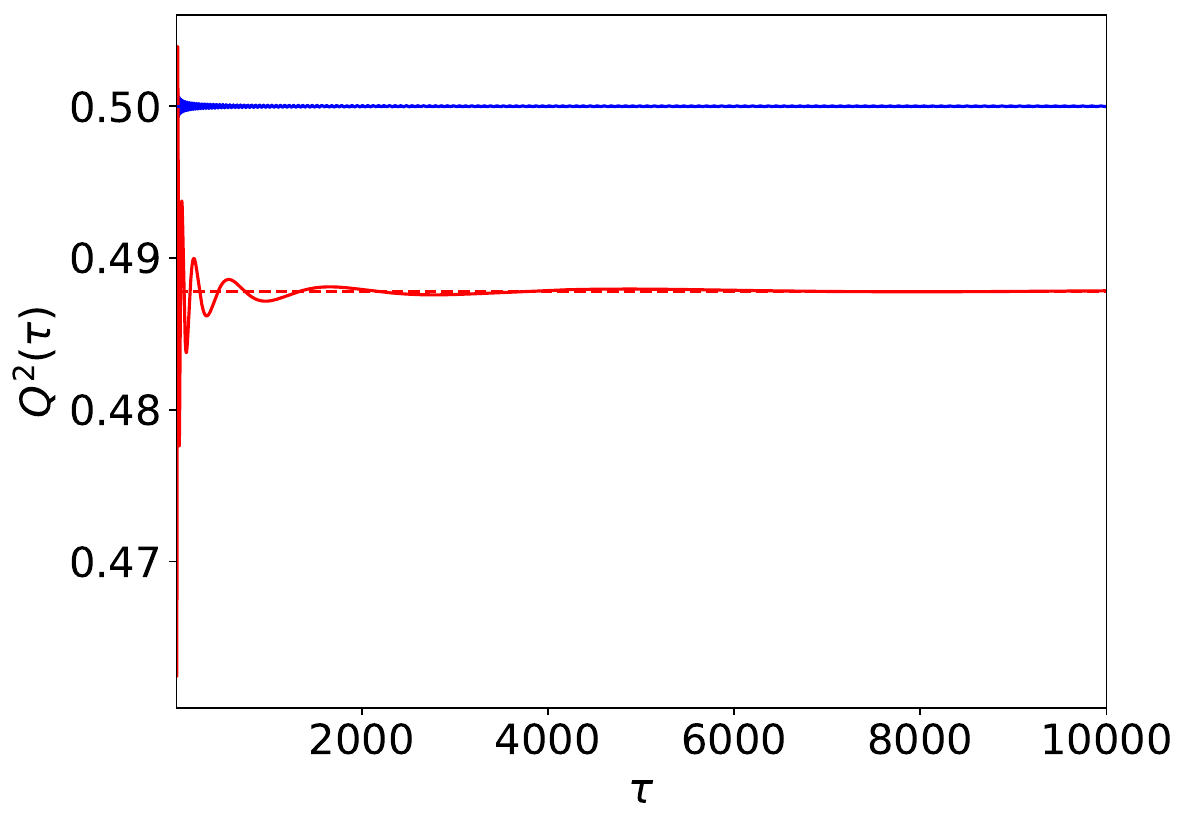}
    \includegraphics[width=0.32\textwidth]{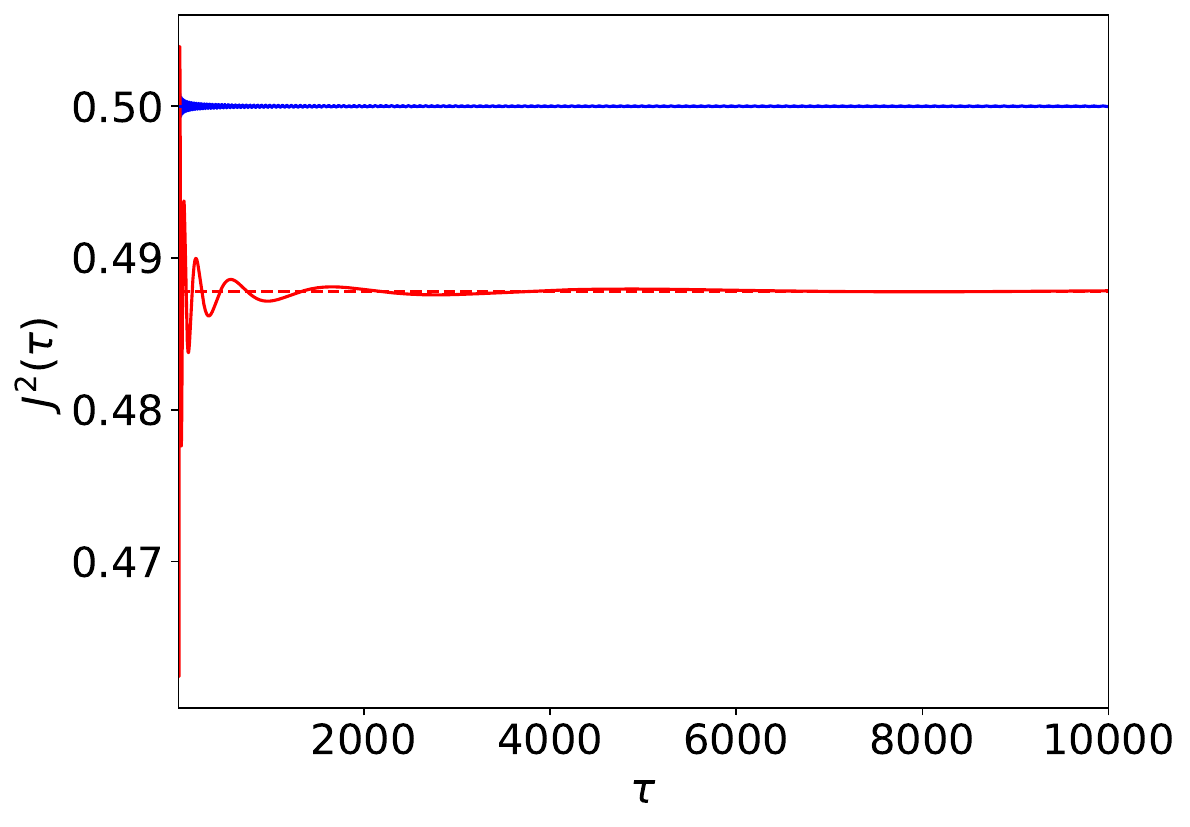}
    \includegraphics[width=0.32\textwidth]{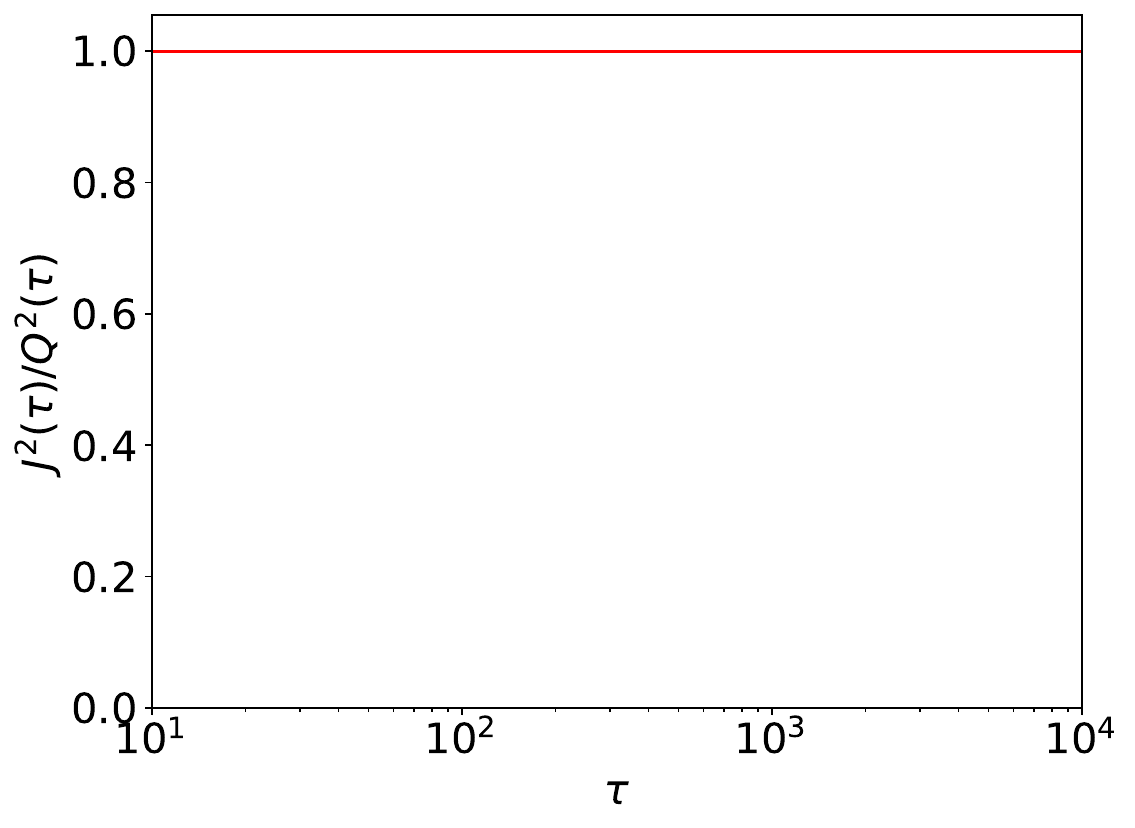}
    \caption{Numerical evolution of networks for the same expansion rate but with different loop production and bias parameters. The two cases correspond to the same initial conditions, but with different values for $\tilde{c}$ and $g$: (Blue) $\tilde{c}=0.1$, $g=0.1$; (Red) $\tilde{c}=0.295$, $g=0.9$.}
    \label{fig03}
\end{figure*}

A final comparison between two qualitatively similar network evolutions, but with different values for $\tilde{c}$ and $g$ is presented in Figure \ref{fig03}, where the distinct scaling of the current and correlation lengths can be easily seen. Note that the selected model parameters have the particularity of ensuring that
\begin{subequations}
\begin{align}
    & \lambda_1 =\frac{2+2\left(1-g_1\right)c_1/k_v}{3+\left(3-2g_1\right)c_1/k_v}\\
    & \lambda_2 = \frac{2-\tilde{c}_2/k_v\left(1-g_2\right)/\mathcal{K}}{1+\tilde{c}_2/k_v}
\end{align}
\end{subequations}
where $\lambda_1$ and $\lambda_2$ are the expansion rates found to be compatible with Eqs. \ref{solutionC2} and Eqs. \ref{solutionC10}, respectively. Assuming $\mathcal{K}=-1/2$ and $k_v$ to be the same in both situations, a suitable choice of $g_i$ and $\tilde{c}_i$
ensures $\lambda_1=\lambda_2$. In other words, this ensures that the same expansion rate will be identifiable with the compatible value with either decaying or constant velocities, depending on the values of the model parameters.

%%%%%%%%%%%%%%%%%%%%%%%%%%%%%%%%%%%%%%%%%%%%%%%%%%%%%%%%%%%%%%%%%%%%%%%%%

%%%%%%%%%%%%%%%%%%%%%%%%%%%%%%%%%%%%%%%%%%%%%%%%%%%%%%%%%%%%%%%%%%%%%%%%%
\section{\label{05conclusions}Discussion and conclusions}

We have now concluded the classification, started in Paper I, of the physically allowed solutions for generic current carrying cosmic string networks, in the framework of the CVOS model. This work extends the analysis previously done for the comparatively simpler cases of wiggly strings \cite{Almeida_2021} and chiral limit strings \cite{Oliveira_2012}. A concise visual comparison of the four classes of solutions is detailed in Table \ref{tab1}.

The first and fourth classes only occur for a single expansion rate. In the case without losses this is $\lambda=2/3$ and $\lambda=2$ (matter era), respectively, but these values shift in the more general case. In the former class the velocities decay as $v\propto a^{-1}$ and the characteristic length scales grow more slowly than the linear scale, while the charge and current remain constant, although in the biased case it is also possible that one of them decays while the other stays constant. The same charge and current behaviour occurs for the latter class, the difference being that in this case the velocity is a constant and the characteristic lengths scale linearly; this is therefore the full scaling solution already mentioned in the introduction.

\begin{table*}
    \centering
    \caption{Comparison of the solutions obtained for different combinations of the phenomenological parameters $\tilde{c}$ and $g$: no losses solutions are obtained for $\tilde{c}=0$; unbiased solutions are obtained for $g=1$, but $\tilde{c}\neq0$; biased solutions are obtained for $\tilde{c}\neq0$ and $g\neq1$. The power law exponents have been expressed with respect to conformal time and comoving length scales. Note that additional consistency conditions apply to these solutions; these have been omitted in the table for simplicity, but are listed in the earlier sections.} 
    \setlength{\extrarowheight}{6pt}
    \resizebox{\textwidth}{!}{
    \begin{tblr}{c c c c c c | c c c c c c | c c c c c c}
    \SetCell[c=6]{c} No losses &&&&&& \SetCell[c=6]{c} Unbiased &&&&&& \SetCell[c=6]{c} Biased &&&&& \\
    $\lambda$ & $\alpha$ & $\beta$ & $\gamma$ & $\delta$ & $\varepsilon$
    & $\lambda$ & $\alpha$ & $\beta$ & $\gamma$ & $\delta$ & $\varepsilon$
    & $\lambda$ & $\alpha$ & $\beta$ & $\gamma$ & $\delta$ & $\varepsilon$
    \\
    \hline
    \SetCell[r=3]{c} $\dfrac{2}{3}$ & \SetCell[r=3]{c} $\dfrac{\lambda}{3}$ & \SetCell[r=3]{c} $-\lambda$ & \SetCell[r=3]{c} 0 & \SetCell[r=3]{c} 0 & \SetCell[r=3]{c} $1-\lambda$ &
    \SetCell[r=3]{c} $\dfrac{2}{3 + \tilde{c}/k_v}$ & \SetCell[r=3]{c} $1-\lambda$ & \SetCell[r=3]{c} $-\lambda$ & \SetCell[r=3]{c} 0 & \SetCell[r=3]{c} 0 & \SetCell[r=3]{c} $1-\lambda$ &
    \SetCell[r=3]{c} $\dfrac{2+2\left(1-g\right)c/k_v}{3 + \left(3-2g\right)\tilde{c}/k_v}$ & \SetCell[r=3]{c} $1-\lambda$ & \SetCell[r=3]{c} $-\lambda$ & 0 & 0 & \SetCell[r=3]{c} $1-\lambda$\\
    &&&&&&&&&&&&&&& 0 & $<0$ &\\
    &&&&&&&&&&&&&&& $<0$ & 0 &\\
    \hline
    $\dfrac{1}{v_0^2}$ & 1 & 0 & $4-2\lambda$ & $4-2\lambda$ & 1
    & $\dfrac{1/v_0^2}{1+\tilde{c}/k_v}$ & 1 & 0 & $4\lambda v_0^2-2\lambda$ & $4\lambda v_0^2-2\lambda$ & 1
    & $<\dfrac{1}{v_0^2 - \mathcal{C}_v\mathcal{K}}\dfrac{1}{1+\tilde{c}/k_v}$ & $\alpha$ & 0 & $2-2\alpha$ & $2-2\alpha$ & 1
    \\
    \hline
    $\dfrac{1}{v_0^2}>2$ & 1 & 0 & $4-2\lambda$ & 0 & 1
    & $\dfrac{1/v_0^2}{1+\tilde{c}/k_v}>\dfrac{2}{1+\tilde{c}/k_v}$ & 1 & 0 & $4\lambda v_0^2-2\lambda$ & 0 & 1
    & \SetCell[r=2]{c} $<\dfrac{2}{1+\tilde{c}/k_v}$ & \SetCell[r=2]{c}{$\alpha$} & \SetCell[r=2]{c}{0} & \SetCell[r=2]{c}{$2-2\alpha$} & \SetCell[r=2]{c}{$2-2\alpha$} & \SetCell[r=2]{c}{1}
    \\
    $\dfrac{1}{v_0^2}>2$ & 1 & 0 & $4-2\lambda$ & $4-2\lambda$ & 1
    & $\dfrac{1/v_0^2}{1+\tilde{c}/k_v}>\dfrac{2}{1+\tilde{c}/k_v}$ & 1 & 0 & $4\lambda v_0^2-2\lambda$ & $4\lambda v_0^2-2\lambda$ & 1
    & & & & & &\\
    \hline
    \SetCell[r=3]{c} 2 & \SetCell[r=3]{c} 1 & \SetCell[r=3]{c} 0 & \SetCell[r=3]{c} 0 & \SetCell[r=3]{c} 0 &
    \SetCell[r=3]{c} 1 & \SetCell[r=3]{c} {$\dfrac{2}{1+\tilde{c}/k_v}$} & \SetCell[r=3]{c} 1 & \SetCell[r=3]{c} 0 &  \SetCell[r=3]{c} 0 & \SetCell[r=3]{c} 0 & 
    \SetCell[r=3]{c} 1 & \SetCell[r=3]{c}{$\dfrac{2-\tilde{c}/k_v\left(1-g\right)/\mathcal{K}}{1+\tilde{c}/k_v}$} &  \SetCell[r=3]{c} 1 &  \SetCell[r=3]{c} 0 & 0 & 0 & \SetCell[r=3]{c} 1 
    \\
    & & & & & 
    & & & & & & 
    & & & & 0 & $<0$ & 
    \\
    & & & & &
    & & & & & &
    & & & & $<0$ & $0$ & 
\end{tblr}}
    \label{tab1}
\end{table*}

On the other hand, the second and third classes are mathematically different but physically similar. The difference is that formally the second class can occur for $v_0^2<1$ while the third can occur for $v_0^2<1/2$, but in practice we expect that the latter condition should apply to all realistic networks. The condition on $\lambda$ for the biased case in the second class may seem different from the others but it is actually similar, bearing in mind that if $g=1$, then $F'=0$ and therefore $\mathcal{K}=0$. These are Nambu-Goto like solutions, with constant velocity and the correlation length $\xi$ scaling linearly. On the other hand, $L$ also scales linearly in the no losses and unbiased cases, though not in the biased case. The behaviour of the charge and current will depend on the values of the parameters (i.e., on whether there are energy losses and/or biases), and the two may evolve differently.

A first overall conclusion is that the additional degrees of freedom in the string worldsheets play a crucial role in the evolution of the networks and this will evidently propagate to the corresponding cosmological consequences. A sufficiently fast expansion rate provides enough damping to make these additional degrees of freedom decay away (even in the absence of further energy loss mechanisms), from which it follows that such networks will asymptotically behave like featureless Nambu-Goto ones. Conversely,  if the damping and energy losses are small, charges and current can dominate the dynamics and prevent the standard and cosmologically benign linear scaling solution. This linear scaling solution does exist (indeed, in an extended version, also including the charge and current themselves, which we have dubbed full scaling), but is confined, for each specific specific choice of model parameters, to a single expansion rate, which in the simplest possible case can be the matter-dominated era.

Two caveats applying to the previous paragraph should nevertheless be kept in mind. The first is that even in the cases in which the Nambu-Goto solution is the mathematical attractor, the timescale needed for the network to effectively reach that limit may be long in cosmological terms---recall that the onset of the matter dominated era took place fairly recently, so even in that case there could be specific astrophysical signatures in the form of comparatively small charges or currents. In such cases, a liner-type model may be adequate \cite{Linear}. The second caveat is that our analysis focused on the evolution of the long strings. Momentarily shifting our attention to loops, these will be considerably smaller than the horizon, and the smaller they are the smaller the effects of the cosmological expansion will be. Therefore on those scales the effects of charges and currents can be more significant, and/or persist for longer. This may impact vorton formation \cite{Martins_1998} as well as the energy decay channels \cite{EMG1,EMg2}. Both of these warrant more in-depth future studies.

A second overall conclusion is that current-carrying string networks form a broad class of models, for which the full range of physical behaviours is, at least in principle, allowed: growing, constant or decaying charge and current, chiral and non-chiral networks, and even solutions with very different charge and current evolutions (e.g., with one of them remaining constant and the other growing or decaying). In different specific models, some of these conceptual possibilities will be actually possible, but not all of them. It is therefore particularly satisfying that the CVOS model explicitly associates the possible behaviours to the microphysics of the model, specifically through the function $f(\kappa)$ whose behaviour is propagated into the macroscopic equation of state $F$ and its first and second derivatives. In this regard, our most noteworthy results are that, according to the CVOS model, biased energy loss and/or non-chiral solutions generally require $\ddprime{F}=0$ and $\dprime{F}\neq0$. Conversely, for $\ddprime{F}\neq0$ and $\dprime{F}=0$ the allowed solutions are more scarce, and often limited to chiral cases.

Several corollaries then stem from these results. Firstly, they confirm that various fundamental and phenomenological particle physics models containing cosmic string networks can be classified according to their cosmological consequences, provided one knows their equation of state, which confirms and contextualizes the early CVOS analysis reported in \cite{Martins_2021}. Secondly, it provides indirect supporting evidence for the CVOS model, in the sense that it shows that the simplifying assumptions and averaging methodology which led to it---and are described in \cite{Martins_2021}---are sufficiently robust to preserve informations on the microphysical specificities of individual models, encoded in $F$ and its derivatives. And thirdly, it suggests that a direct test of the CVOS model, as well as its quantitative calibration, should be feasible for any model whose microphysics can by coded into a field theory simulation.

Until recently, computational bottlenecks related to the use of field-theory based codes prevented, with some exceptions, the simulation of cosmic string networks beyond the featureless Abelian-Higgs ones. The development of highly efficient and scalable GPU-based codes \cite{CUDA1,CUDA2} has removed this bottleneck, and we have recently reported a first implementation in this coding paradigm of the evolution of current-carrying cosmic string networks \cite{Simulations}. Among other things, this work provides tentative evidence for the special role of the matter-dominated era, as well as numerical diagnostics for the coherence lengths of the charge and current. The next logical step is to further exploit this code to carry out extensive series of high-resolution simulations of these models, for several expansion rates, in order to ascertain which of the possible scaling solutions we have described do occur in practice, and to measure, directly from the numerical simulations, the energy loss bias parameters we have explored in this work. This work is currently in progress.

%%%%%%%%%%%%%%%%%%%%%%%%%%%%%%%%%%%%%%%%%%%%%%%%%%%%%%%%%%%%%%%%%%%%%%%%%

\begin{acknowledgments}
This work was financed by Portuguese funds through FCT (Funda\c c\~ao para a Ci\^encia e a Tecnologia) in the framework of the project 2022.04048.PTDC (Phi in the Sky, DOI 10.54499/2022.04048.PTDC). CJM also acknowledges FCT and POCH/FSE (EC) support through Investigador FCT Contract 2021.01214.CEECIND/CP1658/CT0001 (DOI 10.54499/2021.01214.CEECIND/CP1658/CT0001). 
\end{acknowledgments}
 
\bibliography{article}

\begin{thebibliography}{28}%
\makeatletter
\providecommand \@ifxundefined [1]{%
 \@ifx{#1\undefined}
}%
\providecommand \@ifnum [1]{%
 \ifnum #1\expandafter \@firstoftwo
 \else \expandafter \@secondoftwo
 \fi
}%
\providecommand \@ifx [1]{%
 \ifx #1\expandafter \@firstoftwo
 \else \expandafter \@secondoftwo
 \fi
}%
\providecommand \natexlab [1]{#1}%
\providecommand \enquote  [1]{``#1''}%
\providecommand \bibnamefont  [1]{#1}%
\providecommand \bibfnamefont [1]{#1}%
\providecommand \citenamefont [1]{#1}%
\providecommand \href@noop [0]{\@secondoftwo}%
\providecommand \href [0]{\begingroup \@sanitize@url \@href}%
\providecommand \@href[1]{\@@startlink{#1}\@@href}%
\providecommand \@@href[1]{\endgroup#1\@@endlink}%
\providecommand \@sanitize@url [0]{\catcode `\\12\catcode `\$12\catcode
  `\&12\catcode `\#12\catcode `\^12\catcode `\_12\catcode `\%12\relax}%
\providecommand \@@startlink[1]{}%
\providecommand \@@endlink[0]{}%
\providecommand \url  [0]{\begingroup\@sanitize@url \@url }%
\providecommand \@url [1]{\endgroup\@href {#1}{\urlprefix }}%
\providecommand \urlprefix  [0]{URL }%
\providecommand \Eprint [0]{\href }%
\providecommand \doibase [0]{https://doi.org/}%
\providecommand \selectlanguage [0]{\@gobble}%
\providecommand \bibinfo  [0]{\@secondoftwo}%
\providecommand \bibfield  [0]{\@secondoftwo}%
\providecommand \translation [1]{[#1]}%
\providecommand \BibitemOpen [0]{}%
\providecommand \bibitemStop [0]{}%
\providecommand \bibitemNoStop [0]{.\EOS\space}%
\providecommand \EOS [0]{\spacefactor3000\relax}%
\providecommand \BibitemShut  [1]{\csname bibitem#1\endcsname}%
\let\auto@bib@innerbib\@empty
%</preamble>
\bibitem [{\citenamefont {Kibble}(1976)}]{Kibble76}%
  \BibitemOpen
  \bibfield  {author} {\bibinfo {author} {\bibfnamefont {T.~W.~B.}\
  \bibnamefont {Kibble}},\ }\bibfield  {title} {\bibinfo {title} {{Topology of
  Cosmic Domains and Strings}},\ }\href
  {https://doi.org/10.1088/0305-4470/9/8/029} {\bibfield  {journal} {\bibinfo
  {journal} {J. Phys. A}\ }\textbf {\bibinfo {volume} {9}},\ \bibinfo {pages}
  {1387} (\bibinfo {year} {1976})}\BibitemShut {NoStop}%
\bibitem [{\citenamefont {Vilenkin}\ and\ \citenamefont
  {Shellard}(2000)}]{VSbook}%
  \BibitemOpen
  \bibfield  {author} {\bibinfo {author} {\bibfnamefont {A.}~\bibnamefont
  {Vilenkin}}\ and\ \bibinfo {author} {\bibfnamefont {E.~P.~S.}\ \bibnamefont
  {Shellard}},\ }\href@noop {} {\emph {\bibinfo {title} {{Cosmic Strings and
  Other Topological Defects}}}}\ (\bibinfo  {publisher} {Cambridge University
  Press},\ \bibinfo {year} {2000})\BibitemShut {NoStop}%
\bibitem [{\citenamefont {Witten}(1985{\natexlab{a}})}]{Witten_1984}%
  \BibitemOpen
  \bibfield  {author} {\bibinfo {author} {\bibfnamefont {E.}~\bibnamefont
  {Witten}},\ }\bibfield  {title} {\bibinfo {title} {{Superconducting
  Strings}},\ }\href {https://doi.org/10.1016/0550-3213(85)90022-7} {\bibfield
  {journal} {\bibinfo  {journal} {Nucl. Phys. B}\ }\textbf {\bibinfo {volume}
  {249}},\ \bibinfo {pages} {557} (\bibinfo {year}
  {1985}{\natexlab{a}})}\BibitemShut {NoStop}%
\bibitem [{\citenamefont {Witten}(1985{\natexlab{b}})}]{Witten85}%
  \BibitemOpen
  \bibfield  {author} {\bibinfo {author} {\bibfnamefont {E.}~\bibnamefont
  {Witten}},\ }\bibfield  {title} {\bibinfo {title} {{Cosmic Superstrings}},\
  }\href {https://doi.org/10.1016/0370-2693(85)90540-4} {\bibfield  {journal}
  {\bibinfo  {journal} {Phys. Lett. B}\ }\textbf {\bibinfo {volume} {153}},\
  \bibinfo {pages} {243} (\bibinfo {year} {1985}{\natexlab{b}})}\BibitemShut
  {NoStop}%
\bibitem [{\citenamefont {Copeland}\ and\ \citenamefont
  {Kibble}(2010)}]{Copeland_2010}%
  \BibitemOpen
  \bibfield  {author} {\bibinfo {author} {\bibfnamefont {E.~J.}\ \bibnamefont
  {Copeland}}\ and\ \bibinfo {author} {\bibfnamefont {T.~W.~B.}\ \bibnamefont
  {Kibble}},\ }\bibfield  {title} {\bibinfo {title} {{Cosmic Strings and
  Superstrings}},\ }\href {https://doi.org/10.1098/rspa.2009.0591} {\bibfield
  {journal} {\bibinfo  {journal} {Proc. Roy. Soc. Lond. A}\ }\textbf {\bibinfo
  {volume} {466}},\ \bibinfo {pages} {623} (\bibinfo {year} {2010})},\ \Eprint
  {https://arxiv.org/abs/0911.1345} {arXiv:0911.1345 [hep-th]} \BibitemShut
  {NoStop}%
\bibitem [{\citenamefont {Ostriker}\ \emph {et~al.}(1986)\citenamefont
  {Ostriker}, \citenamefont {Thompson},\ and\ \citenamefont {Witten}}]{EMG1}%
  \BibitemOpen
  \bibfield  {author} {\bibinfo {author} {\bibfnamefont {J.~P.}\ \bibnamefont
  {Ostriker}}, \bibinfo {author} {\bibfnamefont {A.~C.}\ \bibnamefont
  {Thompson}},\ and\ \bibinfo {author} {\bibfnamefont {E.}~\bibnamefont
  {Witten}},\ }\bibfield  {title} {\bibinfo {title} {{Cosmological Effects of
  Superconducting Strings}},\ }\href
  {https://doi.org/10.1016/0370-2693(86)90301-1} {\bibfield  {journal}
  {\bibinfo  {journal} {Phys. Lett. B}\ }\textbf {\bibinfo {volume} {180}},\
  \bibinfo {pages} {231} (\bibinfo {year} {1986})}\BibitemShut {NoStop}%
\bibitem [{\citenamefont {Vilenkin}\ and\ \citenamefont
  {Vachaspati}(1987)}]{EMg2}%
  \BibitemOpen
  \bibfield  {author} {\bibinfo {author} {\bibfnamefont {A.}~\bibnamefont
  {Vilenkin}}\ and\ \bibinfo {author} {\bibfnamefont {T.}~\bibnamefont
  {Vachaspati}},\ }\bibfield  {title} {\bibinfo {title} {{Electromagnetic
  Radiation from Superconducting Cosmic Strings}},\ }\href
  {https://doi.org/10.1103/PhysRevLett.58.1041} {\bibfield  {journal} {\bibinfo
   {journal} {Phys. Rev. Lett.}\ }\textbf {\bibinfo {volume} {58}},\ \bibinfo
  {pages} {1041} (\bibinfo {year} {1987})}\BibitemShut {NoStop}%
\bibitem [{\citenamefont {Saffin}(2005)}]{Saffin05}%
  \BibitemOpen
  \bibfield  {author} {\bibinfo {author} {\bibfnamefont {P.~M.}\ \bibnamefont
  {Saffin}},\ }\bibfield  {title} {\bibinfo {title} {{A Practical model for
  cosmic (p,q) superstrings}},\ }\href
  {https://doi.org/10.1088/1126-6708/2005/09/011} {\bibfield  {journal}
  {\bibinfo  {journal} {JHEP}\ }\textbf {\bibinfo {volume} {09}},\ \bibinfo
  {pages} {011}},\ \Eprint {https://arxiv.org/abs/hep-th/0506138}
  {arXiv:hep-th/0506138} \BibitemShut {NoStop}%
\bibitem [{\citenamefont {Urrestilla}\ and\ \citenamefont
  {Vilenkin}(2008)}]{Urrestilla_2008}%
  \BibitemOpen
  \bibfield  {author} {\bibinfo {author} {\bibfnamefont {J.}~\bibnamefont
  {Urrestilla}}\ and\ \bibinfo {author} {\bibfnamefont {A.}~\bibnamefont
  {Vilenkin}},\ }\bibfield  {title} {\bibinfo {title} {{Evolution of cosmic
  superstring networks: A Numerical simulation}},\ }\href
  {https://doi.org/10.1088/1126-6708/2008/02/037} {\bibfield  {journal}
  {\bibinfo  {journal} {JHEP}\ }\textbf {\bibinfo {volume} {02}},\ \bibinfo
  {pages} {037}},\ \Eprint {https://arxiv.org/abs/0712.1146} {arXiv:0712.1146
  [hep-th]} \BibitemShut {NoStop}%
\bibitem [{\citenamefont {Lizarraga}\ and\ \citenamefont
  {Urrestilla}(2016)}]{Lizarraga16}%
  \BibitemOpen
  \bibfield  {author} {\bibinfo {author} {\bibfnamefont {J.}~\bibnamefont
  {Lizarraga}}\ and\ \bibinfo {author} {\bibfnamefont {J.}~\bibnamefont
  {Urrestilla}},\ }\bibfield  {title} {\bibinfo {title} {{Survival of
  pq-superstrings in field theory simulations}},\ }\href
  {https://doi.org/10.1088/1475-7516/2016/04/053} {\bibfield  {journal}
  {\bibinfo  {journal} {JCAP}\ }\textbf {\bibinfo {volume} {04}},\ \bibinfo
  {pages} {053}},\ \Eprint {https://arxiv.org/abs/1602.08014} {arXiv:1602.08014
  [astro-ph.CO]} \BibitemShut {NoStop}%
\bibitem [{\citenamefont {Correia}\ and\ \citenamefont
  {Martins}(2022)}]{Correia_2022}%
  \BibitemOpen
  \bibfield  {author} {\bibinfo {author} {\bibfnamefont {J.~R. C. C.~C.}\
  \bibnamefont {Correia}}\ and\ \bibinfo {author} {\bibfnamefont {C.~J. A.~P.}\
  \bibnamefont {Martins}},\ }\bibfield  {title} {\bibinfo {title} {Multitension
  strings in high-resolution
  $\mathrm{U}(1)\ifmmode\times\else\texttimes\fi{}\mathrm{U}(1)$ simulations},\
  }\href {https://doi.org/10.1103/PhysRevD.106.043521} {\bibfield  {journal}
  {\bibinfo  {journal} {Phys. Rev. D}\ }\textbf {\bibinfo {volume} {106}},\
  \bibinfo {pages} {043521} (\bibinfo {year} {2022})}\BibitemShut {NoStop}%
\bibitem [{\citenamefont {Battye}\ and\ \citenamefont
  {Cotterill}(2023)}]{Battye23}%
  \BibitemOpen
  \bibfield  {author} {\bibinfo {author} {\bibfnamefont {R.~A.}\ \bibnamefont
  {Battye}}\ and\ \bibinfo {author} {\bibfnamefont {S.~J.}\ \bibnamefont
  {Cotterill}},\ }\bibfield  {title} {\bibinfo {title} {{Pinching instabilities
  in superconducting cosmic strings}},\ }\href
  {https://doi.org/10.1103/PhysRevD.107.063534} {\bibfield  {journal} {\bibinfo
   {journal} {Phys. Rev. D}\ }\textbf {\bibinfo {volume} {107}},\ \bibinfo
  {pages} {063534} (\bibinfo {year} {2023})},\ \Eprint
  {https://arxiv.org/abs/2212.06491} {arXiv:2212.06491 [hep-ph]} \BibitemShut
  {NoStop}%
\bibitem [{\citenamefont {Correia}\ \emph {et~al.}(2024)\citenamefont
  {Correia}, \citenamefont {Martins},\ and\ \citenamefont
  {Pimenta}}]{Simulations}%
  \BibitemOpen
  \bibfield  {author} {\bibinfo {author} {\bibfnamefont {J.~R. C. C.~C.}\
  \bibnamefont {Correia}}, \bibinfo {author} {\bibfnamefont {C.~J. A.~P.}\
  \bibnamefont {Martins}},\ and\ \bibinfo {author} {\bibfnamefont {F.~C.
  N.~Q.}\ \bibnamefont {Pimenta}},\ }\bibfield  {title} {\bibinfo {title}
  {{Evolution of current-carrying string networks}},\ }\href
  {https://doi.org/10.1016/j.physletb.2024.138788} {\bibfield  {journal}
  {\bibinfo  {journal} {Phys. Lett. B}\ }\textbf {\bibinfo {volume} {855}},\
  \bibinfo {pages} {138788} (\bibinfo {year} {2024})},\ \Eprint
  {https://arxiv.org/abs/2406.03931} {arXiv:2406.03931 [hep-ph]} \BibitemShut
  {NoStop}%
\bibitem [{\citenamefont {Martins}\ and\ \citenamefont
  {Shellard}(1996)}]{VOS1}%
  \BibitemOpen
  \bibfield  {author} {\bibinfo {author} {\bibfnamefont {C.~J. A.~P.}\
  \bibnamefont {Martins}}\ and\ \bibinfo {author} {\bibfnamefont {E.~P.~S.}\
  \bibnamefont {Shellard}},\ }\bibfield  {title} {\bibinfo {title}
  {{Quantitative string evolution}},\ }\href
  {https://doi.org/10.1103/PhysRevD.54.2535} {\bibfield  {journal} {\bibinfo
  {journal} {Phys. Rev. D}\ }\textbf {\bibinfo {volume} {54}},\ \bibinfo
  {pages} {2535} (\bibinfo {year} {1996})},\ \Eprint
  {https://arxiv.org/abs/hep-ph/9602271} {arXiv:hep-ph/9602271} \BibitemShut
  {NoStop}%
\bibitem [{\citenamefont {Martins}\ and\ \citenamefont
  {Shellard}(2002)}]{Martins_2002}%
  \BibitemOpen
  \bibfield  {author} {\bibinfo {author} {\bibfnamefont {C.~J. A.~P.}\
  \bibnamefont {Martins}}\ and\ \bibinfo {author} {\bibfnamefont {E.~P.~S.}\
  \bibnamefont {Shellard}},\ }\bibfield  {title} {\bibinfo {title} {Extending
  the velocity-dependent one-scale string evolution model},\ }\href
  {https://doi.org/10.1103/PhysRevD.65.043514} {\bibfield  {journal} {\bibinfo
  {journal} {Phys. Rev. D}\ }\textbf {\bibinfo {volume} {65}},\ \bibinfo
  {pages} {043514} (\bibinfo {year} {2002})}\BibitemShut {NoStop}%
\bibitem [{\citenamefont {Martins}(2016)}]{Martins_2016}%
  \BibitemOpen
  \bibfield  {author} {\bibinfo {author} {\bibfnamefont {C.}~\bibnamefont
  {Martins}},\ }\href
  {https://doi.org/https://doi.org/10.1007/978-3-319-44553-3} {\emph {\bibinfo
  {title} {Defect Evolution in Cosmology and Condensed Matter: Quantitative
  Analysis with the Velocity-Dependent One-Scale Model}}},\ \bibinfo {edition}
  {1st}\ ed.\ (\bibinfo  {publisher} {Springer Cham},\ \bibinfo {year}
  {2016})\BibitemShut {NoStop}%
\bibitem [{\citenamefont {Martins}\ \emph
  {et~al.}(2021{\natexlab{a}})\citenamefont {Martins}, \citenamefont {Peter},
  \citenamefont {Rybak},\ and\ \citenamefont {Shellard}}]{Martins_2021}%
  \BibitemOpen
  \bibfield  {author} {\bibinfo {author} {\bibfnamefont {C.~J. A.~P.}\
  \bibnamefont {Martins}}, \bibinfo {author} {\bibfnamefont {P.}~\bibnamefont
  {Peter}}, \bibinfo {author} {\bibfnamefont {I.~Y.}\ \bibnamefont {Rybak}},\
  and\ \bibinfo {author} {\bibfnamefont {E.~P.~S.}\ \bibnamefont {Shellard}},\
  }\bibfield  {title} {\bibinfo {title} {{Generalized velocity-dependent
  one-scale model for current-carrying strings}},\ }\href
  {https://doi.org/10.1103/PhysRevD.103.043538} {\bibfield  {journal} {\bibinfo
   {journal} {Phys. Rev. D}\ }\textbf {\bibinfo {volume} {103}},\ \bibinfo
  {pages} {043538} (\bibinfo {year} {2021}{\natexlab{a}})},\ \Eprint
  {https://arxiv.org/abs/2011.09700} {arXiv:2011.09700 [astro-ph.CO]}
  \BibitemShut {NoStop}%
\bibitem [{\citenamefont {Rybak}\ \emph {et~al.}(2023)\citenamefont {Rybak},
  \citenamefont {Martins}, \citenamefont {Peter},\ and\ \citenamefont
  {Shellard}}]{Rybak_2023}%
  \BibitemOpen
  \bibfield  {author} {\bibinfo {author} {\bibfnamefont {I.~Y.}\ \bibnamefont
  {Rybak}}, \bibinfo {author} {\bibfnamefont {C.~J. A.~P.}\ \bibnamefont
  {Martins}}, \bibinfo {author} {\bibfnamefont {P.}~\bibnamefont {Peter}},\
  and\ \bibinfo {author} {\bibfnamefont {E.~P.~S.}\ \bibnamefont {Shellard}},\
  }\bibfield  {title} {\bibinfo {title} {{Cosmological evolution of Witten
  superconducting string networks}},\ }\href
  {https://doi.org/10.1103/PhysRevD.107.123514} {\bibfield  {journal} {\bibinfo
   {journal} {Phys. Rev. D}\ }\textbf {\bibinfo {volume} {107}},\ \bibinfo
  {pages} {123514} (\bibinfo {year} {2023})},\ \Eprint
  {https://arxiv.org/abs/2304.00053} {arXiv:2304.00053 [astro-ph.CO]}
  \BibitemShut {NoStop}%
\bibitem [{\citenamefont {Correia}\ and\ \citenamefont
  {Martins}(2021{\natexlab{a}})}]{Correia21}%
  \BibitemOpen
  \bibfield  {author} {\bibinfo {author} {\bibfnamefont {J.~R. C. C.~C.}\
  \bibnamefont {Correia}}\ and\ \bibinfo {author} {\bibfnamefont {C.~J. A.~P.}\
  \bibnamefont {Martins}},\ }\bibfield  {title} {\bibinfo {title} {{High
  resolution calibration of the cosmic strings velocity dependent one-scale
  model}},\ }\href {https://doi.org/10.1103/PhysRevD.104.063511} {\bibfield
  {journal} {\bibinfo  {journal} {Phys. Rev. D}\ }\textbf {\bibinfo {volume}
  {104}},\ \bibinfo {pages} {063511} (\bibinfo {year} {2021}{\natexlab{a}})},\
  \Eprint {https://arxiv.org/abs/2108.07513} {arXiv:2108.07513 [astro-ph.CO]}
  \BibitemShut {NoStop}%
\bibitem [{\citenamefont {Pimenta}\ and\ \citenamefont
  {Martins}(2024)}]{PaperI}%
  \BibitemOpen
  \bibfield  {author} {\bibinfo {author} {\bibfnamefont {F.~C. N.~Q.}\
  \bibnamefont {Pimenta}}\ and\ \bibinfo {author} {\bibfnamefont {C.~J. A.~P.}\
  \bibnamefont {Martins}},\ }\bibfield  {title} {\bibinfo {title} {{Scaling
  solutions for current-carrying cosmic string networks}},\ }\href
  {https://doi.org/10.1103/PhysRevD.110.023540} {\bibfield  {journal} {\bibinfo
   {journal} {Phys. Rev. D}\ }\textbf {\bibinfo {volume} {110}},\ \bibinfo
  {pages} {023540} (\bibinfo {year} {2024})},\ \Eprint
  {https://arxiv.org/abs/2406.03940} {arXiv:2406.03940 [hep-ph]} \BibitemShut
  {NoStop}%
\bibitem [{\citenamefont {Oliveira}\ \emph {et~al.}(2012)\citenamefont
  {Oliveira}, \citenamefont {Avgoustidis},\ and\ \citenamefont
  {Martins}}]{Oliveira_2012}%
  \BibitemOpen
  \bibfield  {author} {\bibinfo {author} {\bibfnamefont {M.~F.}\ \bibnamefont
  {Oliveira}}, \bibinfo {author} {\bibfnamefont {A.}~\bibnamefont
  {Avgoustidis}},\ and\ \bibinfo {author} {\bibfnamefont {C.~J. A.~P.}\
  \bibnamefont {Martins}},\ }\bibfield  {title} {\bibinfo {title} {Cosmic
  string evolution with a conserved charge},\ }\href@noop {} {\bibfield
  {journal} {\bibinfo  {journal} {Physical Review D}\ }\textbf {\bibinfo
  {volume} {85}},\ \bibinfo {pages} {083515} (\bibinfo {year}
  {2012})}\BibitemShut {NoStop}%
\bibitem [{\citenamefont {Almeida}\ and\ \citenamefont
  {Martins}(2021)}]{Almeida_2021}%
  \BibitemOpen
  \bibfield  {author} {\bibinfo {author} {\bibfnamefont {A.~R.~R.}\
  \bibnamefont {Almeida}}\ and\ \bibinfo {author} {\bibfnamefont {C.~J. A.~P.}\
  \bibnamefont {Martins}},\ }\bibfield  {title} {\bibinfo {title} {Scaling
  solutions of wiggly cosmic strings},\ }\href
  {https://doi.org/10.1103/PhysRevD.104.043524} {\bibfield  {journal} {\bibinfo
   {journal} {Phys. Rev. D}\ }\textbf {\bibinfo {volume} {104}},\ \bibinfo
  {pages} {043524} (\bibinfo {year} {2021})}\BibitemShut {NoStop}%
\bibitem [{\citenamefont {Almeida}\ and\ \citenamefont
  {Martins}(2022)}]{Almeida_2022}%
  \BibitemOpen
  \bibfield  {author} {\bibinfo {author} {\bibfnamefont {A.~R.~R.}\
  \bibnamefont {Almeida}}\ and\ \bibinfo {author} {\bibfnamefont {C.~J. A.~P.}\
  \bibnamefont {Martins}},\ }\bibfield  {title} {\bibinfo {title} {Scaling
  solutions of wiggly cosmic strings. ii. time-varying coarse-graining scale
  solutions},\ }\href {https://doi.org/10.1103/PhysRevD.106.083525} {\bibfield
  {journal} {\bibinfo  {journal} {Phys. Rev. D}\ }\textbf {\bibinfo {volume}
  {106}},\ \bibinfo {pages} {083525} (\bibinfo {year} {2022})}\BibitemShut
  {NoStop}%
\bibitem [{\citenamefont {Pimenta}(2023)}]{Thesis}%
  \BibitemOpen
  \bibfield  {author} {\bibinfo {author} {\bibfnamefont {F.~C. N.~Q.}\
  \bibnamefont {Pimenta}},\ }\emph {\bibinfo {title} {{Analytical solutions for
  the evolution of current carrying cosmic strings}}},\ \href@noop {} {Master's
  thesis},\ \bibinfo  {school} {University of Porto} (\bibinfo {year}
  {2023})\BibitemShut {NoStop}%
\bibitem [{\citenamefont {Martins}\ \emph
  {et~al.}(2021{\natexlab{b}})\citenamefont {Martins}, \citenamefont {Peter},
  \citenamefont {Rybak},\ and\ \citenamefont {Shellard}}]{Linear}%
  \BibitemOpen
  \bibfield  {author} {\bibinfo {author} {\bibfnamefont {C.~J. A.~P.}\
  \bibnamefont {Martins}}, \bibinfo {author} {\bibfnamefont {P.}~\bibnamefont
  {Peter}}, \bibinfo {author} {\bibfnamefont {I.~Y.}\ \bibnamefont {Rybak}},\
  and\ \bibinfo {author} {\bibfnamefont {E.~P.~S.}\ \bibnamefont {Shellard}},\
  }\bibfield  {title} {\bibinfo {title} {{Charge-velocity-dependent one-scale
  linear model}},\ }\href {https://doi.org/10.1103/PhysRevD.104.103506}
  {\bibfield  {journal} {\bibinfo  {journal} {Phys. Rev. D}\ }\textbf {\bibinfo
  {volume} {104}},\ \bibinfo {pages} {103506} (\bibinfo {year}
  {2021}{\natexlab{b}})},\ \Eprint {https://arxiv.org/abs/2108.03147}
  {arXiv:2108.03147 [astro-ph.CO]} \BibitemShut {NoStop}%
\bibitem [{\citenamefont {Martins}\ and\ \citenamefont
  {Shellard}(1998)}]{Martins_1998}%
  \BibitemOpen
  \bibfield  {author} {\bibinfo {author} {\bibfnamefont {C.~J. A.~P.}\
  \bibnamefont {Martins}}\ and\ \bibinfo {author} {\bibfnamefont {E.~P.~S.}\
  \bibnamefont {Shellard}},\ }\bibfield  {title} {\bibinfo {title} {Vorton
  formation},\ }\href {https://doi.org/10.1103/PhysRevD.57.7155} {\bibfield
  {journal} {\bibinfo  {journal} {Phys. Rev. D}\ }\textbf {\bibinfo {volume}
  {57}},\ \bibinfo {pages} {7155} (\bibinfo {year} {1998})}\BibitemShut
  {NoStop}%
\bibitem [{\citenamefont {Correia}\ and\ \citenamefont
  {Martins}(2020)}]{CUDA1}%
  \BibitemOpen
  \bibfield  {author} {\bibinfo {author} {\bibfnamefont {J.~R. C. C.~C.}\
  \bibnamefont {Correia}}\ and\ \bibinfo {author} {\bibfnamefont {C.~J. A.~P.}\
  \bibnamefont {Martins}},\ }\bibfield  {title} {\bibinfo {title}
  {{Abelian-Higgs Cosmic String Evolution with CUDA}},\ }\href
  {https://doi.org/10.1016/j.ascom.2020.100388} {\bibfield  {journal} {\bibinfo
   {journal} {Astron. Comput.}\ }\textbf {\bibinfo {volume} {32}},\ \bibinfo
  {pages} {100388} (\bibinfo {year} {2020})},\ \Eprint
  {https://arxiv.org/abs/1809.00995} {arXiv:1809.00995 [physics.comp-ph]}
  \BibitemShut {NoStop}%
\bibitem [{\citenamefont {Correia}\ and\ \citenamefont
  {Martins}(2021{\natexlab{b}})}]{CUDA2}%
  \BibitemOpen
  \bibfield  {author} {\bibinfo {author} {\bibfnamefont {J.~R. C. C.~C.}\
  \bibnamefont {Correia}}\ and\ \bibinfo {author} {\bibfnamefont {C.~J. A.~P.}\
  \bibnamefont {Martins}},\ }\bibfield  {title} {\bibinfo {title}
  {{Abelian\textendash{}Higgs cosmic string evolution with multiple GPUs}},\
  }\href {https://doi.org/10.1016/j.ascom.2020.100438} {\bibfield  {journal}
  {\bibinfo  {journal} {Astron. Comput.}\ }\textbf {\bibinfo {volume} {34}},\
  \bibinfo {pages} {100438} (\bibinfo {year} {2021}{\natexlab{b}})},\ \Eprint
  {https://arxiv.org/abs/2005.14454} {arXiv:2005.14454 [physics.comp-ph]}
  \BibitemShut {NoStop}%
\end{thebibliography}%
\end{document}